\documentclass[1p,final]{elsarticle}

\usepackage{graphics}
\usepackage{graphicx}
\usepackage{epsfig}

\usepackage{lineno}

\usepackage{amssymb}

\begin{document}


\newcommand{\microg}{$\mu$g}
\newcommand{\microm}{$\mu$g}
\newcommand{\heavywater}{${\rm D}_2{\rm O}$}
\newcommand{\water}{${\rm H}_2{\rm O}$}
\newcommand{\methane}{${\rm CH}_4$}
\newcommand{\isotope}[2]{$^{#2}{\rm #1}$}

\begin{frontmatter}


\title{Four methods for determining the composition of trace radioactive surface contamination of low-radioactivity metal \\ 
}

\author[Oxford]{H.\,M.\,O'Keeffe\fnref{HelenNow}}
\author[CENPA]{T.\,H.\,Burritt}
\author[SNOLab]{B.\,T.\,Cleveland}
\author[Oxford]{G.\,Doucas}
\author[SNOLab]{N.\,Gagnon}
\author[Oxford]{N.\,A.\,Jelley}
\author[Laurentian]{C.\,Kraus}
\author[SNOLab]{I.\,T.\,Lawson}
\author[Oxford]{S.\,Majerus}
\author[CENPA]{S.\,R.\,McGee\fnref{SeanNow}}
\author[CENPA]{A.\,W.\,Myers}
\author[LBNL]{A.\,W.\,P.\,Poon}
\author[LANL,CENPA]{K.\,Rielage}
\author[CENPA]{R.\,G.\,H.\,Robertson\corref{Hamish}}
\author[CENPA]{R.\,C.\,Rosten}
\author[LANL,CENPA]{L.\,C.\,Stonehill}
\author[CENPA]{B.\,A.\,VanDevender\fnref{BrentNow}}
\author[CENPA]{T.\,D.\,Van Wechel}

\address[Oxford]{Department of Physics, University of Oxford, Denys Wilkinson Building, Oxford, OX1 3RH, UK}
\address[LANL]{Los Alamos National Laboratory, Los Alamos, NM 87545, USA}
\address[CENPA]{Center for Experimental Nuclear Physics and Astrophysics, and Department of Physics, University of Washington, Seattle, WA 98195, USA}
\address[SNOLab]{SNOLAB, Lively, Ontario, P3Y 1M3, Canada}
\address[LBNL]{Institute for Nuclear and Particle Astrophysics, and Nuclear Science Division, Lawrence Berkeley National Laboratory, Berkeley, CA 94720, USA}
\address[Laurentian]{Department of Physics and Astronomy, Laurentian University, Sudbury, Ontario P3E 2C6, Canada}

\fntext[HelenNow]{Present Address: Department of Physics, Queen's University, Kingston, Ontario K7L 3N6, Canada}
\fntext[SeanNow]{Present Address: Department of Genome Sciences, University of Washington, Seattle, WA, 98195, USA}
\fntext[BrentNow]{Present Address: Pacific Northwest National Laboratory, Richland, WA 99352, USA}

\cortext[Hamish]{Corresponding author.  rghr@u.washington.edu}


\begin{abstract}
Four methods for determining the composition of low-level uranium- and thorium-chain surface contamination are presented.   One method is the observation of Cherenkov light production in water.  In two additional methods a  position-sensitive proportional counter surrounding the surface is used to make both a measurement of the energy spectrum of alpha particle emissions and also coincidence measurements to derive the thorium-chain content based on the presence of short-lived isotopes in that decay chain.  The fourth method is a radiochemical technique in which the surface is eluted with a weak acid, the eluate is concentrated, added to liquid scintillator and assayed by recording beta-alpha coincidences.   These methods were used to characterize two `hotspots' on the outer surface of one of the $^3$He proportional counters in the Neutral Current Detection array of the Sudbury Neutrino Observatory experiment.  The methods have similar sensitivities, of order tens of ng,  to both thorium- and uranium-chain contamination.
 \end{abstract}


\end{frontmatter}


\section{Introduction}\label{sec:intro}

Trace amounts of \isotope{U}{238}, \isotope{Th}{232}, and their daughters, referred to collectively as the uranium and thorium chains, are pervasive in common materials, and their presence plagues low-background experiments such as the Sudbury Neutrino Observatory (SNO).  SNO was a 1\,000-tonne heavy-water (\heavywater) Cherenkov detector viewed by 9\,456 photomultiplier tubes (PMTs) situated 2\,072 m below the surface in Vale INCO's Creighton mine in Sudbury, Ontario, Canada~\cite{snonim}.  Neutrinos participate primarily in three interactions in \heavywater~\cite{HChen}.  The interaction of most interest for this work is the neutral current (NC) interaction, in which neutrinos of any flavor can dissociate a deuteron into its constituent proton and neutron ($\nu_x \ + \ {\rm D} \to \nu_x \ + \ {\rm p \ + \ n}$).  The detection of the neutron constitutes SNO's signal of a NC event.  Uranium- and thorium-chain contamination is a problem for SNO because near the end of each chain gammas are produced  with sufficient energy to photodisintegrate a deuteron in the \heavywater \ target volume.  This results in a neutron that cannot be distinguished from a neutron created by a NC interaction, and thus represents a background.    The branching ratio and interaction cross section for the  thorium-chain gamma are larger than for uranium, making thorium-chain contamination the more severe problem for SNO.  Strategies  to mitigate the effects are to limit contamination through the use of ultra-clean materials and purification techniques, and to determine precisely how much remains so that the resulting background can be calculated and subtracted from the data.  

\begin{figure}
\begin{center}\includegraphics[width=.7\textwidth]{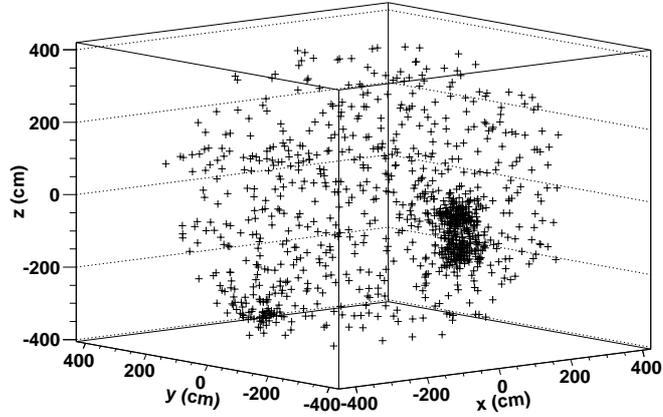}\end{center}
\caption{A scatter plot of the $(x,y,z)$ coordinates of reconstructed PMT array vertices in a selected low-energy event window.  Regions of elevated activity are clearly visible.  Events in the pair of larger active spots  at $(x,y,z) \approx (50, -200, -120)$ and $\approx (50, -200, -30)$\,cm are clustered about the NCD K5.  The smaller spot [$(x,y,z) \approx (-250, 50, -300)$\,cm] corresponds to the position of another NCD, K2.  That spot was analyzed similarly to the K5 hotspots discussed here.}
\label{fig:xyz}
\end{figure}

Even the most careful handling of materials can result in unexpected contamination, causing problematic backgrounds for sensitive experiments.  Three areas of increased radioactivity associated with the SNO Neutral Current Detection  array were identified.  The array~\cite{ncdnim} consisted of 36 strings of \isotope{He}{3}-filled proportional counters (hereafter referred to as `NCDs') to detect neutrons liberated by NC interactions, plus 4 additional \isotope{He}{4}-filled strings to give a sample of non-neutron backgrounds.  The NCD active volumes were enclosed in cylindrical nickel shells, 5\,cm in diameter with lengths from 9--11\,m.  The array was installed into the \heavywater \ for the third phase of SNO operations to give a measure of the NC flux independent of the charged-current and elastic-scattering fluxes measured by the PMT array.  The vertices of reconstructed low-energy events in the fiducial volume were disproportionately concentrated at 3 locations near the NCDs ``K2'' and ``K5,''  as shown in Fig.~\ref{fig:xyz}.  These regions will be termed `hotspots'.
\begin{figure}
\begin{center}\includegraphics[width=.7\textwidth]{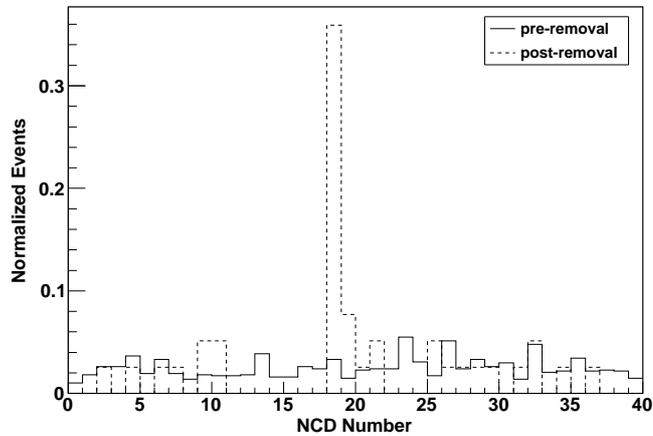}\end{center}
\caption{Comparison of 72 hours of data taken before the removal of K5 (``pre-removal'') with 72 hours of data taken afterwards (``post-removal'').  The histogram shows the relative event rates in a low-energy background window with vertices occurring inside a 50\,cm radius of each NCD, and also within a total fiducal volume defined by a cylinder with axis from $z = \pm 250$\,cm and radius $r_{xy} = 350$\,cm.  K5 is NCD number 18.  The hotspot is clearly visible after removal, but not before.  A slightly higher overall count rate in the pre-removal data  was observed and is attributed to elevated levels of $^{222}$Rn which were present due to the absence of cover gas during array installation.  The hotspot would have been visible above these levels, if it had been present at that time.}
\label{fig:beforeAndAfter}
\end{figure}
K5 is unique among the NCDs because after initial installation in the \heavywater \ it had to be raised partway up the neck of the acrylic vessel to diagnose problems unrelated to contamination.  The problems persisted after it was restored to the array and the data from K5 were not used.
Figure~\ref{fig:beforeAndAfter} shows PMT data taken in two equal time intervals, one before K5's removal and one after its reinstallation.  The activity  on K5 is concentrated in two spots (Fig.~\ref{fig:K5z}).  The positions of the spots are in the general vicinity of where the surface of the water would have contacted the string for a time during the raising operation.
It is clear that K5 was somehow  contaminated during its removal and reinstallation, but it is unknown how it happened.  
\begin{figure}
\begin{center}\includegraphics[width=.7\textwidth]{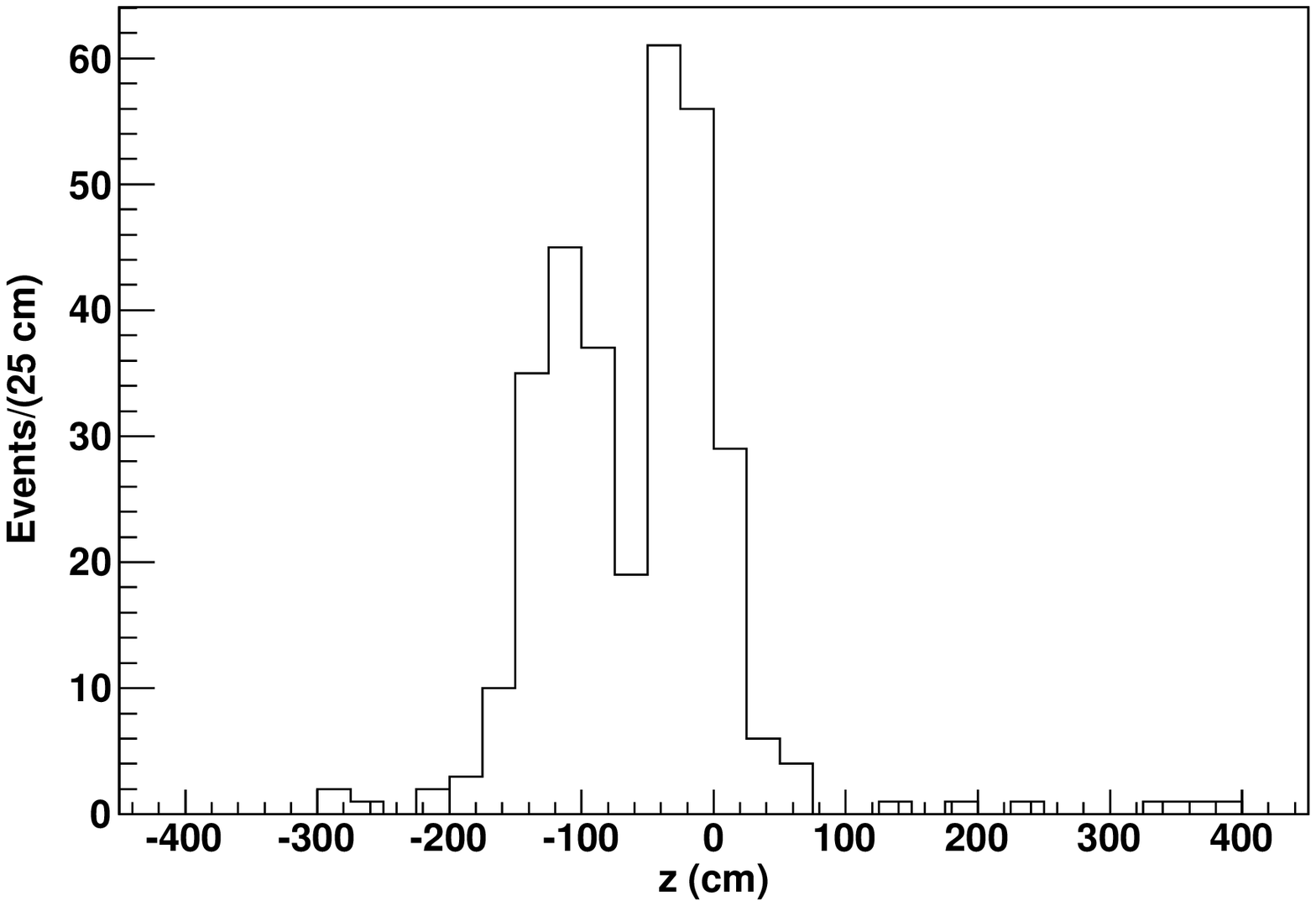}\end{center}
\caption{The $z$ coordinate of events that reconstruct nearest to K5.  In SNO, the axes of the NCDs are parallel to $z$.  There appear to be two active spots.  The width of the peaks is dominated by the resolution of the vertex reconstruction.  These are the same events as those in bin 18 of Figure~\ref{fig:beforeAndAfter}.}
\label{fig:K5z}
\end{figure}

A preliminary analysis showed that if the composition of these hotspots remained unknown, the resulting uncertainty in backgrounds could negatively affect the accuracy of the NC flux measurement~\cite{ncdprl}, adding an additional 4\,\% systematic uncertainty to the anticipated 7--8\,\% total uncertainty budget. An analysis of the events observed {\em in-situ} by the PMT array could not exclude the possibility that the excess count rate was caused by something other than radioactivity. For  example, the increased activity could have been due to a deposit on the surface of the NCD capable of scintillating.  This would have increased the signal strength of any small amount of radioactivity, without creating additional background.  Radioactivity could produce neutrons, so an attempt was made to search for excess neutrons.  The majority of neutrons should capture back on the contaminated counter~\cite{PMT}.  Unfortunately, K5 suffered from operational problems mentioned above, and this analysis proved inconclusive.

Knowledge of the composition and strength of the activity could reduce the systematic uncertainty in the NC flux, so a number of tests were conducted on K5 after it was removed at the end of the experiment.  The deck area above SNO's heavy-water vessel is maintained as a clean room of class 1000 or better, with strict procedures in place to preserve cleanliness~\cite{snonim}.  K5 was removed from the \heavywater \ directly into this environment and immediately stored under a tent to preclude any additional contamination that might have been introduced by traffic in and out of the deck clean room, which was kept to the absolute minimum required to analyze the hotspots.  K5 was not handled in the regions near the suspected contamination.  In addition to the PMT analysis (referred to as the Cherenkov method), three independent methods were developed to measure the surface activity on K5.  An External Alpha Counter (EAC) was built by the University of Washington for the first two methods.  The EAC is a position-sensitive multi-wire proportional flow chamber.  Such a device can have much better position resolution than the PMT event vertex reconstruction algorithm used to make the plot in Figure~\ref{fig:K5z}.    Alpha spectra from the EAC give detailed information  about the intensities at each step in a decay chain (EAC spectral-analysis method) and the same instrument can be used to count time-correlated alpha events   (EAC coincidence method). The fourth technique (Acid-elution method) was developed by the University of Oxford. After counting in the EAC, the surface material in the contaminated regions of the NCD detector were chemically removed  and concentrated in solution with liquid scintillator.  Time-correlated beta-alpha decay pairs were counted.

Each method exploits features of the uranium and thorium decay chains to obtain good determination of the contamination composition with well-established techniques and analyses.   Disequilibrium within the chains can occur at the time of contamination because of the different affinities of the elements involved for the nickel surface of the NCD.  In addition, with a surface deposit, an alpha decay can eject the daughter.  The result is a disturbance of the relative abundances of decay-chain isotopes.  As will be shown, the initial contamination appears to have been highly selective in depositing thorium isotopes.  The uranium chain is not observed at any significant level,  including even the thorium isotope 75-ky \isotope{Th}{230}.  In the Th chain, 1.91-y \isotope{Th}{228} is found to be present, but apparently unsupported by its parent 5.75-y   \isotope{Ra}{228}.   Leaching of Ra, the most probable explanation, would affect  \isotope{Ra}{228} and 1600-y \isotope{Ra}{226} similarly, effectively terminating the U chain at \isotope{Ra}{226}.    The gammas that trouble SNO come from the decays of \isotope{Bi}{214} and \isotope{Tl}{208} in the lower parts of the uranium and thorium chains, respectively.

In our EAC spectral analysis method, we measure the total alpha spectrum in the EAC and fit it to a set of lines corresponding to each alpha decay.   The line strength is expressed in terms of the equivalent equilibrium amounts of \isotope{U}{238} and \isotope{Th}{232}  in the hotspots.  The energy spectrum of thorium-chain alphas has two ``gaps'' around 4.5\,MeV and 8\,MeV, while uranium-chain alphas tend to occupy those same regions, providing good sensitivity for the presence of both chains.

The EAC coincidence method searches the same EAC data for the time-correlated alpha decay events in the series \isotope{Ra}{224} $\rightarrow$ \isotope{Rn}{220} $\rightarrow$ \isotope{Po}{216} $\rightarrow$ \isotope{Pb}{212} in the lower thorium chain.  The half-lives of $^{220}$Rn and $^{216}$Po are only 56\,s and  0.145\,s, respectively.    The probability of observing a complete triple coincidence is reduced relative to that of a double coincidence.  After accounting for detection efficiencies, the number of double coincidences relates to the lower-thorium-chain content in the same way as the parameters of the spectral analysis, but in a spectrum that is much less complex.

The Acid-elution method is also based on time-correlated decays  in the lower uranium and thorium decay chains.  The series \isotope{Bi}{212} $\rightarrow$ \isotope{Po}{212} $\rightarrow$ \isotope{Pb}{208} occurs at the end of 64\,\% of all thorium decay chains.  The leading beta decay is followed shortly by the trailing alpha decay, due to the 0.3-$\mu$s half-life of \isotope{Po}{212}.   The uranium decay chain contains a similar beta-alpha coincidence, \isotope{Bi}{214} $\rightarrow$ \isotope{Po}{214} $\rightarrow$ \isotope{Pb}{210}, correlated in time by the 164-$\mu$s half-life of \isotope{Po}{214}.  The large disparity between the two polonium half-lives can be used to distinguish lower uranium-chain activity from lower thorium-chain activity.     Pulse-shape analysis of scintillation light can discriminate between  betas and alphas  (see e.\,g.,~\cite{Knoll}).   We therefore removed the activity from the NCD surface by eluting it with a weak acid, concentrated the resulting solution, added it to a small volume of liquid scintillator, and observed the scintillation light  with a single PMT.  The number of observed coincidences relates to the amounts of  \isotope{U}{238} and \isotope{Th}{228} as in the other methods.

All four methods were successfully applied to the K5 hotspots.  The Cherenkov method has excellent signal-to-background and located the activity with modest spatial resolution.  The two alpha-counting methods located the activity on two spots, each a few cm long.   The $\alpha$-spectral analysis determined the contribution of the upper and lower parts of both the uranium and thorium chains to the activity of the hotspots.  It is the only one of our methods with sensitivity to the upper chains.  The EAC coincidence method corroborates the EAC spectral analysis results for the lower thorium chain.  The Acid-elution method determined the contribution of the lower parts of the uranium and thorium chains.  The methods generally agree within experimental uncertainty.  Knowledge of the composition reduced the potential 4\,\% systematic uncertainty discussed above to $<$1\,\% by allowing a more precise prediction of the number of background neutrons created by the K5 hotspots.   For brevity, we omit detailed discussion of the K2 hotspot, but include the results at the end.

\section{Cherenkov analysis}
The Cherenkov method measured the hotspot radioactivity using data taken 
with the PMT array during the third phase of SNO operation, 2004--2006.  Analysis cuts were applied to the data to select the 
background dominated region.  The cuts defined the Cherenkov analysis 
window, between 4.5 and 5.0 MeV total energy, in which the selected events 
should be dominated by the resolution tails of  $^{214}$Bi and $^{208}$Tl from the uranium and 
thorium chains respectively.

The amount of each was determined by exploiting differences in the pattern of
hit PMTs produced when Cherenkov light created by decays was detected.  Single-electron events formed a Cherenkov cone that, although broadened by multiple 
scattering in the water, could be characterized by a relatively tight cluster
of PMT hits.  Multiple-electron events consist of a convolution of several single 
electron events with different directional distributions and therefore, on average, a more 
isotropic distribution of hit PMTs.  The event-isotropy parameter \cite{Ahmad:2002jz} could 
therefore be used to distinguish between single- and multiple-cone events, which are characteristic of  $^{214}$Bi and  $^{208}$Tl, respectively. 

Beta emission from $^{208}$Tl leads  to excited states of 
$^{208}$Pb that decay by  gamma emission through the 
2.614-MeV state.  The $^{214}$Bi decays, on the other hand, are dominated 
by the 18\% direct beta transition to the ground state of $^{214}$Po, with energetic gammas being emitted in less than 3\% of decays.  Both the isotropy and the light yield therefore depend on whether the activity is dissolved in the water, deposited on the surface of an NCD, or embedded within an NCD.  Monte Carlo methods were used to simulate the observable effects for these possibilities.
The mean isotropy of K5 events was found to be consistent with a combination of surface
$^{208}$Tl and $^{214}$Bi.  A 1-dimensional extended maximum 
likelihood fit (see \cite{PDG}) was made using surface isotropy distributions.  The contributions of  
D$_2$O and NCD bulk backgrounds determined by other analyses were constrained in the fit.  The
mass of each contaminant based on the extracted surface activities is given in 
Table \ref{ism}.  
\begin{table}[!h]
\caption{Results from the Cherenkov analysis of the two K5 hotspots together.
Total uncertainties are given.  \label{ism} }
\begin{center}
\begin{tabular}{c c}
\hline 
\vspace{-0.1in} &  \\
$^{232}$Th & $1.48^{+0.24}_{-0.27}$ $\mu$g\\
\vspace{-0.1in} &  \\
$^{238}$U  & $0.77^{+0.19}_{-0.23}$ $\mu$g \\
\vspace{-0.1in} &  \\
\hline
\end{tabular}
\end{center}
\end{table}

While the Cherenkov analysis was consistent with surface
$^{208}$Tl and $^{214}$Bi, a variety of other conceivable light-producing contaminants cannot be ruled out by this method alone.  Methods that provide complementary and specific information about the composition of the hotspots are
discussed in the remainder of this paper.

\section{The External Alpha Counter}\label{sec:EAC}

\begin{figure}[ht]
\begin{center}\includegraphics[width=.67\textwidth]{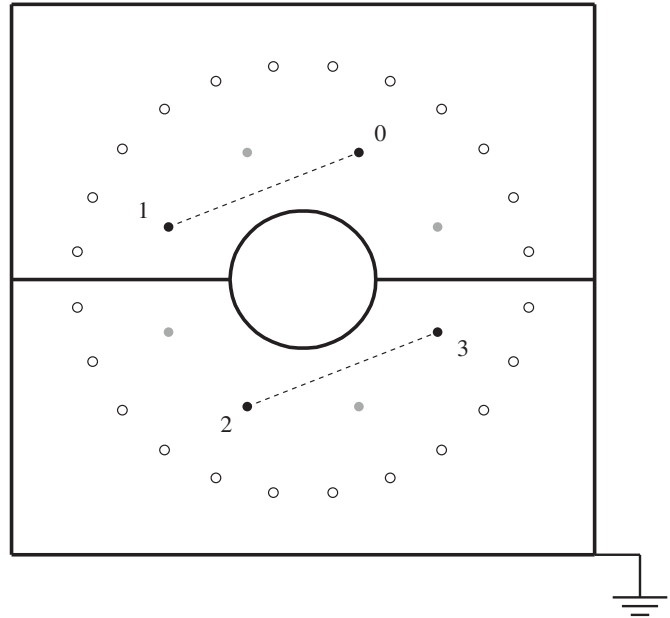}\end{center}
\caption{A diagram of the EAC, viewed from the electronics end.  It has the form of a rectangular box split into two halves that can be separated for easy placement of an NCD into the hole.  There is a ring of 8 resistive wires, concentric with the NCD.  The dotted lines show how the active anodes 0--3 are connected together in alternating pairs at the far end.  The other 4 wires in the inner ring are at ground potential ($\equiv 0$\,V).  The outer ring of wires is held at a small negative potential to reject ionization events coming from outside the gas region occupied by the NCD.  }
\label{fig:EACdiagram}
\end{figure}

The EAC, sketched in Figure~\ref{fig:EACdiagram}, is a multi-wire proportional flow counter.
\begin{figure}[ht]
\begin{center}\includegraphics[width=.67\textwidth]{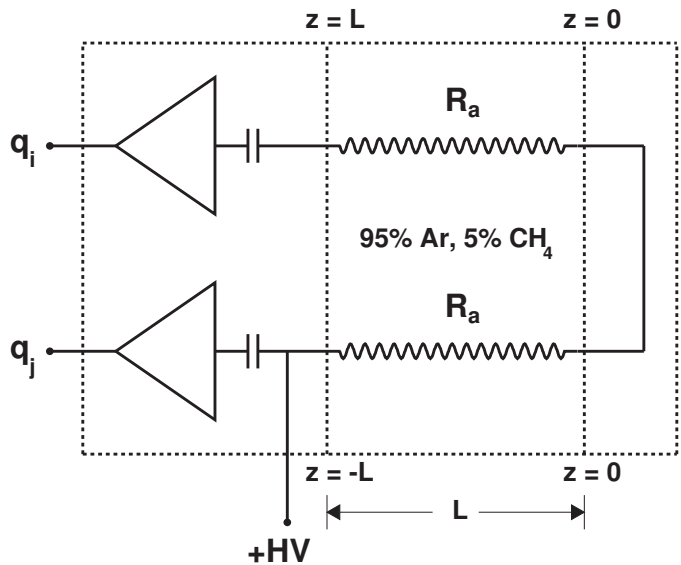}\end{center}
\caption{The charge-division circuit of the EAC.  Resistive anodes with resistance $R_a$ and length $L$ are grouped in pairs connected at the far end, and the collected charge $q$ is read out at the electronics end.  The charge-division algorithm, Eq.~(\ref{eqn:chargeDiv}), determines which anode segment actually collected the charge at position $z$ in each event.}
\label{fig:circuitDiagram}
\end{figure}
It is designed to be sensitive to low levels of radioactivity, while being simple to construct and operate.  The symmetric halves can be easily separated for quick insertion and removal of a 5-cm diameter NCD.  When assembled, the counter completely encloses a 122-cm length of the NCD.  The outer surface of the enclosed length of NCD is open to the active gas-filled region of the EAC.  The total length of the EAC includes 8\,cm in the ``electronics end'' which is filled with wire-binding posts, preamplifiers, and signal and high-voltage connections.  Another 5\,cm at the ``far end'' contains binding posts and  jumpers that connect alternate pairs of anode wires.  The remaining 109\,cm is strung with wire and filled with P5 gas (95\,\% Ar, 5\,\% \methane).  The three regions are separated from each other by stainless steel walls welded into an outer stainless-steel frame.  No special low-radioactivity materials were used, with the exception of non-thoriated welding rods and electrodes, and copper wire and mounting terminations.  The two halves of the EAC fit closely around the NCD, but no elastomeric seals are used to contain the gas since the distribution of radioactivity was not known before counting and seals could remove (or add) surface contamination.  P5 flows into the active region at 1 standard m$^3$ per hour to keep the active volume slightly above ambient pressure and prevent radon-rich air in the underground lab from entering the chamber.  The NCD is surrounded by a concentric 10-cm diameter ring of 8 25-$\mu$m diameter Stablohm resistive wires running parallel to the NCD.    They are crimped inside 1.5-mm-diameter copper tubes that extend approximately 4\,cm into the active volume from each end, thus reducing the length of the position-sensitive volume to 101\,cm and excluding alphas from the end walls.   The gas gain is optimized in a configuration where these wires alternate between high-voltage anode and ground potential, as shown in Figure~\ref{fig:EACdiagram}.  Concentric with the NCD and the anode ring is a passive ring with diameter 17\,cm consisting of 24 50-$\mu$m diameter copper wires held at a small negative potential ($V_g = -100\,{\rm V}$), to suppress events occurring in the outer volume of the EAC.

Using resistive anode wires with preamps at each end, the longitudinal location of events can be determined by a simple charge-division algorithm.  Grouping the anodes in pairs and collecting charge at only one end of the counter eliminates the ground loop that would arise if the charge were collected at both ends of the counter (see Figures~\ref{fig:EACdiagram} and~\ref{fig:circuitDiagram}).  Most of the electrons liberated by an alpha emitted from the surface of the NCD will collect at a single anode, with only $\sim 5$\% of events resulting in charge on neighboring anodes.  This is expected, given the range of alphas and the $90^\circ$ separation between active anodes.  From Figure~\ref{fig:circuitDiagram} we see that if the electrons from an ionization event are collected at position $z_i$ along anode $i$, the charges $q$ collected at the ends of anodes $i$ and $j$ are related to $z_i$ by
\begin{equation}\label{eqn:chargeDiv}
z =  \frac{ q_i- q_j}{q_i + q_j}  L, \  \left\{
\begin{array}{rl}
z \geq 0: & {\rm anode \ } i \\
z < 0: & {\rm anode \ } j, z \rightarrow -z \\
\end{array}
\right.
\end{equation}

\section{EAC Spectral Analysis}\label{sec:spectralAnalysis}

\subsection{Event Selection and Background Subtraction}

\begin{figure}
\includegraphics[width=0.5\textwidth]{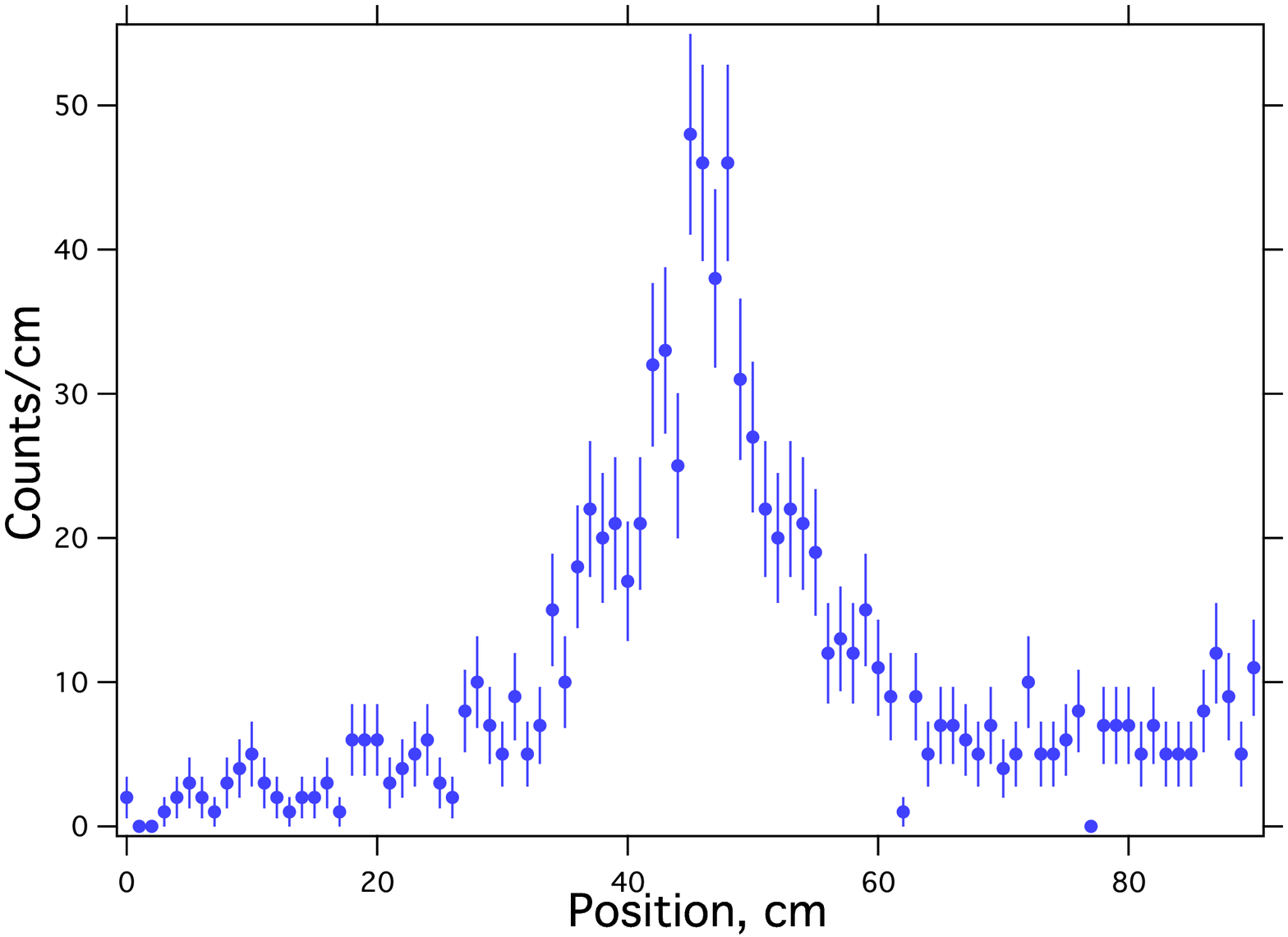}
\includegraphics[width=0.5\textwidth]{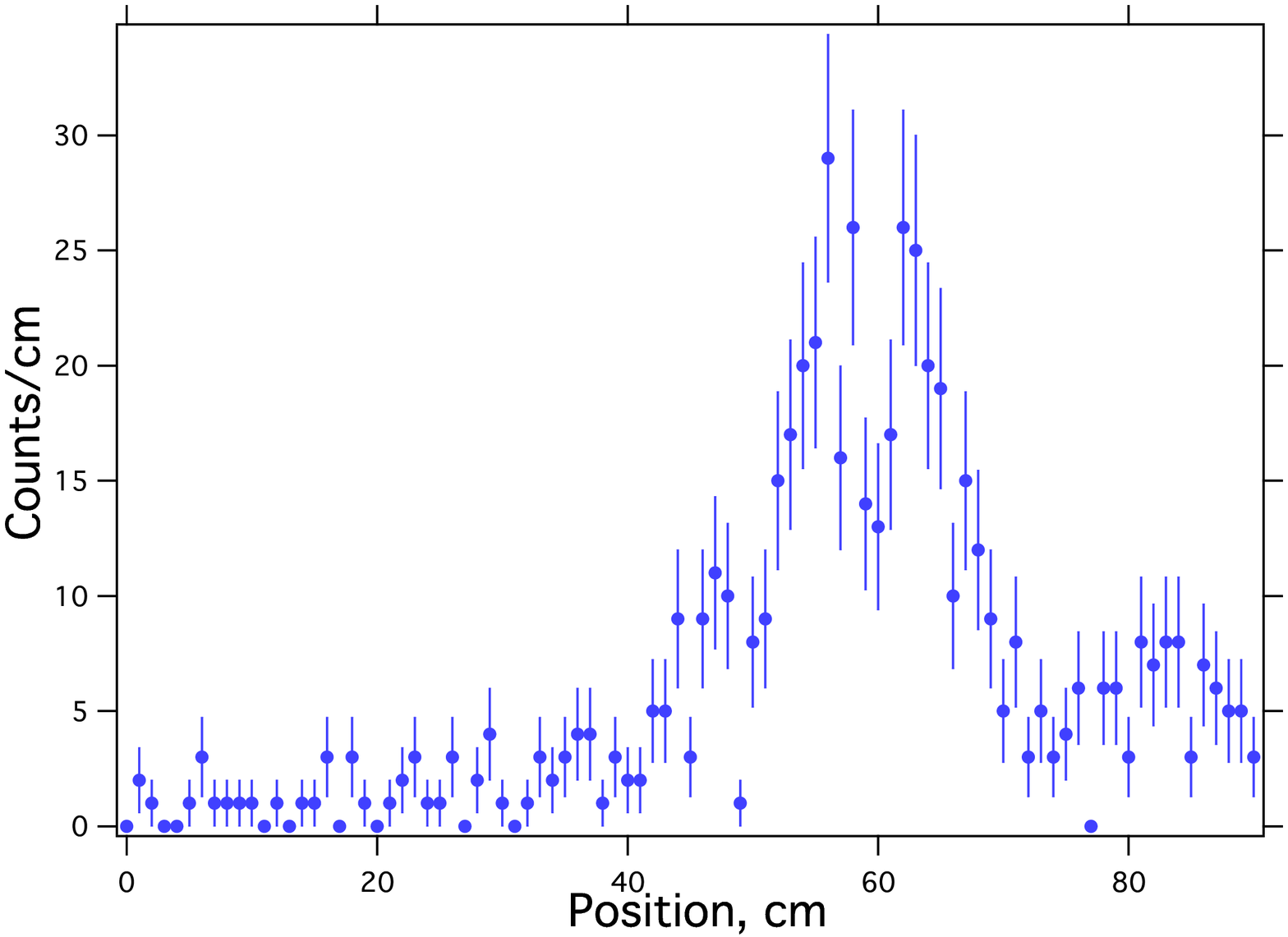}
\begin{center}
\caption{\label{fig:Z} The $z$-positions of events recorded by the EAC for: (left panel) the upper K5 hotspot, and (right panel) the lower K5 hotspot.    Each plot represents approximately 24 hours of data.}
\end{center}
\end{figure}

K5 was counted in two passes, one for each spot.  It was placed in the EAC such that the centers of the hotspot distributions, as determined from the Cherenkov analysis (Figure~\ref{fig:K5z}), were approximately centered in the active volume with the minimum between peaks at one end.  
A section of uncontaminated NCD (from the NCD labeled K2) was also counted to establish the background due to \isotope{Po}{210}, which is uniformly distributed on the surfaces of all NCDs~\cite{ncdnim}, as well as instrumental backgrounds inherent to the EAC.   
The $z$ positions of events in the K5 hotspot data sets are shown in Figure~\ref{fig:Z}.  
A distribution of events with about 20\,cm full width at half maximum (FWHM) is clearly visible above the background in both sets.  The background is due  to \isotope{Po}{210}, which is uniformly distributed on the surfaces of all NCDs~\cite{ncdnim}, and \isotope{Rn}{220} and \isotope{Rn}{222}, which are present in the gas.  A peak at 100\,cm due to high-voltage microdischarges at the wire supports of the EAC is cut off in this display.

\subsection{Data Analysis}

\begin{figure}
\includegraphics[width=0.5\textwidth]{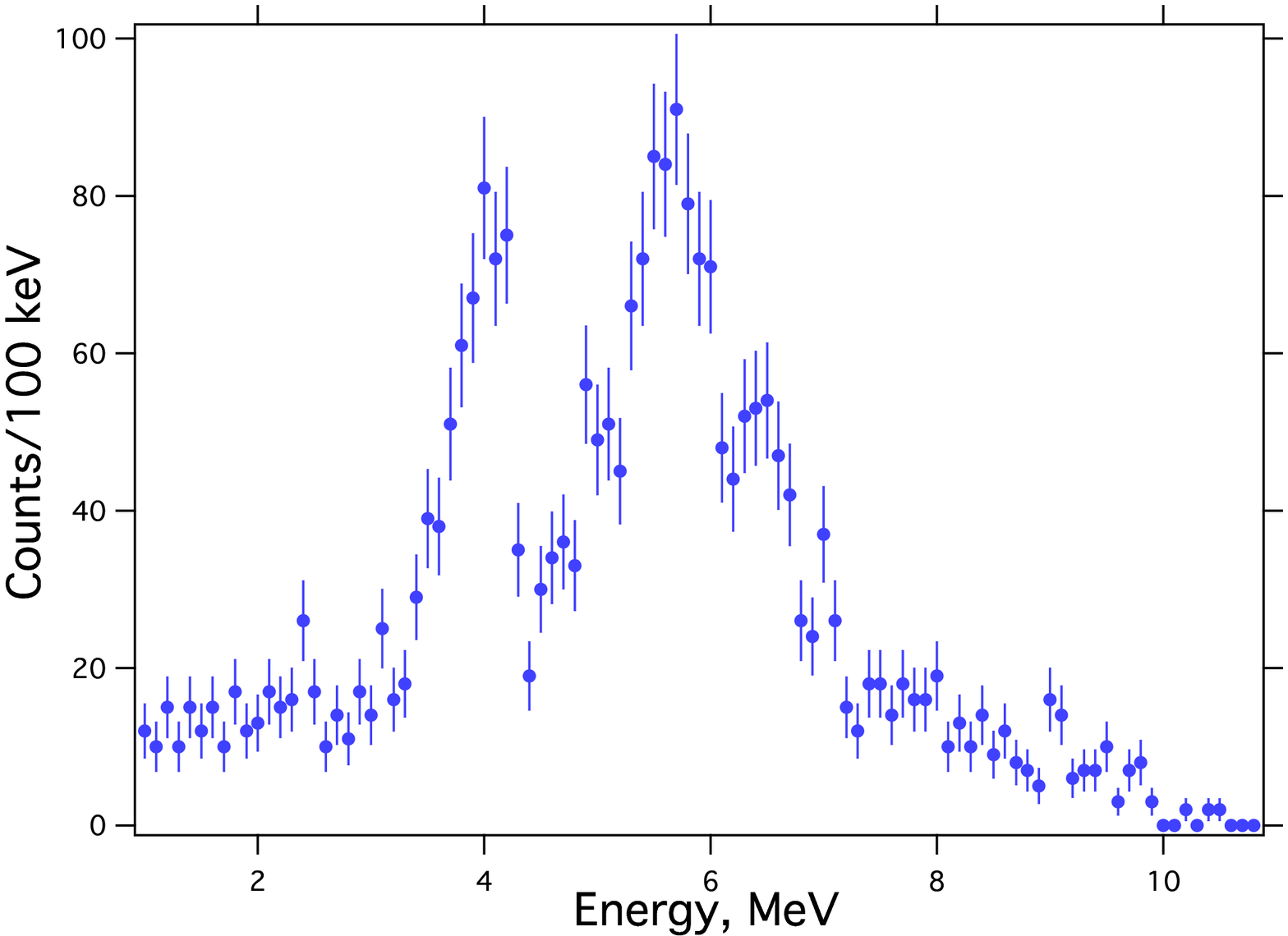}
\includegraphics[width=0.5\textwidth]{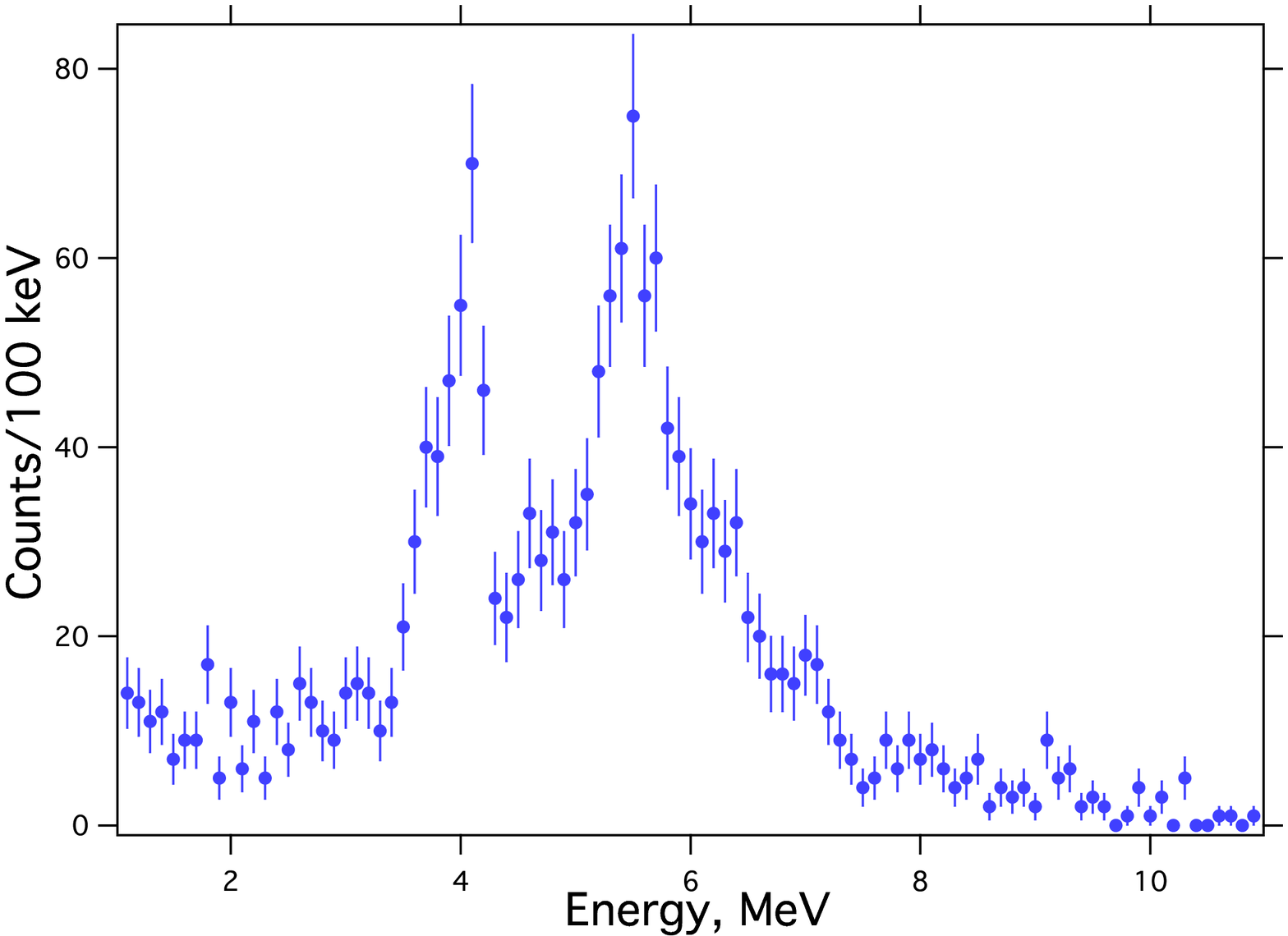}
\begin{center}
\caption{\label{fig:E} The alpha-energy spectra of: (left panel) the upper K5 hotspot, and (right panel) the lower K5 hotspot.  }
\end{center}
\end{figure}

An energy calibration of the EAC was made with 59.45-keV gammas from an \isotope{Am}{241} source.  However, nonlinearities limit the accuracy of extrapolations up to alpha energies $\sim$ 1--10\,MeV.  Furthermore, the gain of the EAC fluctuates with the ambient pressure in the underground lab by up to $\pm\!$ 2\,\% from day to day.  So, while the \isotope{Am}{241} calibration is sufficient for qualitative identification of spectral features, fine tuning of the calibration is required for rigorous quantitative analysis.  The  data from both hotspots shows a clear peak around 4.0\,MeV, corresponding to alphas of a single energy (actually a close doublet) from the decay of \isotope{Th}{232} in the upper thorium chain.  Energy-calibration points for the EAC can be obtained internally from each data set by using this 4-MeV line and two isolated lines at 6.288 MeV  and  6.778 MeV from \isotope{Rn}{220} and  \isotope{Po}{216} in the Th series obtained by imposing a time-coincidence requirement of 0.3 s.    Random coincidences and coincidences from the \isotope{Ra}{224}-\isotope{Rn}{220} sequence are subtracted. A linear calibration fits the 3 lines well.  The background arising from \isotope{Rn}{220} and \isotope{Rn}{222} in the gas and the surface-deposited \isotope{Po}{210} is removed by setting a position window away from the hotspot and subtracting that spectrum.  For the upper hotspot, the signal and background regions chosen were 33 to 60 cm and 65 to 85 cm, respectively.  The signal region for the lower hotspot was chosen to be 45 to 76 cm, while the background was taken from two regions, 25 to 40 and 75 to 85 cm.
The energy spectrum of events in the signal region is shown for each hotspot in Figure~\ref{fig:E}, 
and background-subtracted spectra in Figure~\ref{fig:Fit}.

Nearly Gaussian line shapes fit the energy spectra, indicating that the deposit is mainly confined to the surface. Fits are improved, however, with the addition of a small exponential tail to give a modified line shape:  
\begin{eqnarray*}
g(E,E_\alpha,\sigma)& = &\exp{\left[\frac{-(E-E_{\alpha})^2}{2\sigma^2}\right]} +m\exp{\frac{E-E_\alpha}{c}}\left(1+\exp{\frac{E-E_\alpha}{\sigma}}\right)^{-1},
\end{eqnarray*}
where $E_\alpha$ is the energy of an alpha line, and $\sigma$, $m$, and $c$ are parameters defining the line shape $g(E,E_\alpha, \sigma)$.  The area under the line is 
\begin{eqnarray*}
G(\sigma,m,c)& = & \int_{0}^{\infty}g(E,E_{\alpha},\sigma) dE \\
&\simeq& \int_{-\infty}^{+\infty}g(E,E_{\alpha},\sigma) dE \\
& = & \sigma\sqrt{2\pi} + mc\Gamma{\left(1 + \frac{\sigma}{c}\right)}\Gamma{\left(1-\frac{\sigma}{c}\right)}
\end{eqnarray*}
for $E_\alpha \gg c >\sigma $.  With data binned in steps of width $\Delta E$ the number of events in a line is
\begin{eqnarray*}
 n & \simeq  &\frac{A}{\Delta E} G(\sigma,m,c) 
\end{eqnarray*}
where $A$ is the fitted amplitude.  In many cases, lines are doublets or multiplets, and can be fitted with a single normalization if the branching ratios $b_j$ are known:
\begin{eqnarray*}
n_{ij}(E_{\alpha,ij})& \simeq  &\frac{A_i b_j}{\Delta E} G(\sigma_{ij},m,c).
\end{eqnarray*}
 The index $i$ denotes each group of related transitions, and the index $j$ denotes a transition within the group. Groupings are based on consideration of chemical and physical factors that might lead to disequilibrium.  The branching ratios $b_j$ are listed in Table~\ref{tab:chains}. 

The width parameter $\sigma$ is assumed to scale as a power $k$ of the energy, representing the range of different physical mechanisms for line broadening that may be present:
\begin{eqnarray*}
\sigma_{ij} &=& \sigma_4\left(\frac{E_{\alpha,ij}}{E_4}\right)^k,
\end{eqnarray*}
where $\sigma_4$ is the line width at a standard energy, $E_4 = 4$ MeV.
  The value of $k$ was determined from fits to the 4-MeV $^{232}$Th line and the two lines at 6.288 and 6.778 MeV extracted from coincidence data.  There was no evidence for an energy dependence: $k=-0.02 \pm 0.05$.  For subsequent analysis, $k=0$ was used.  The fitted parameter values for the K5 upper and lower hotspots are listed in Table~\ref{tab:parameters}.

  \begin{table}
\caption{Th and U decay-chain groupings for independent intensity fits for the K5 upper and lower hotspots.  The energies and intensities are derived from \cite{Isotopes}.}
\medskip 
\begin{center}
\begin{tabular}{cccrr}
\hline
 Isotope & Energy (MeV) & Branch & \multicolumn{2}{c}{Mass $m_i$, $\mu$g}  \\
 \multicolumn{3}{c}{}&K5 Upper& K5 Lower \\
 \hline
\multicolumn{5}{l}{\bf Th Chain}  \\
\hline
 $^{232}$Th & 4.012 & 0.782  & 2.7(7) &1.8(2) \\
$^{232}$Th & 3.947 & 0.217 & \\
\hline
$^{228}$Th & 5.423 & 0.722 &  0.8(4) & 0.3(3) \\
$^{228}$Th & 5.340 & 0.272 & \\
\hline
$^{224}$Ra & 5.680 & 1& 1.6(5) & 1.2(3) \\
\hline
$^{220}$Rn & 6.288 & 1&  0.7(2) & 0.6(1) \\
$^{216}$Po & 6.778 & 1 &    &  \\
\hline
$^{212}$Bi & 6.090 & 0.097 & 0.3(1) & 0.1(1)\\
$^{212}$Bi & 6.051 & 0.251 &  & \\
$^{212}$Po & 8.784 & 0.641&  &  \\
\hline
\multicolumn{5}{l}{\bf U Chain}  \\
\hline
 $^{238}$U & 4.198 & 0.790 & $\leq 0.12$  & $\leq 0.08$ \\
$^{238}$U & 4.151 & 0.209 & \\
$^{234}$U & 4.775 & 0.714 & \\
$^{234}$U & 4.722 & 0.284 & \\
\hline
$^{230}$Th & 4.687 & 0.763 & 0.01(7) & 0.12(6) \\
$^{230}$Th & 4.621 & 0.234 &  \\
\hline
$^{226}$Ra & 4.780 & 1 &  $\equiv 0$  & $\equiv 0$ \\
\hline
$^{222}$Rn & 5.590 & 1& bkg.  \\
\hline
$^{218}$Po & 6.002 & 1 & 0.05(4) & 0.06(2)  \\
$^{214}$Po & 7.687 & 1 &  \\
\hline
$^{210}$Po & 5.304 & 1 & bkg.  \\
\hline
\end{tabular}
\end{center}
\label{tab:chains}
\end{table}

  \begin{table}
\caption{Parameters used in or determined from fits to singles and coincidence data.}
\medskip 
\begin{center}
\begin{tabular}{lccl}
\hline
Parameter   & K5 Upper & K5 Lower  & Units \\
 \hline
S($^{232}$Th) & \multicolumn{2}{c}{4079.6} & Bq/g \\
S($^{238}$U) & \multicolumn{2}{c}{12455.2} & Bq/g \\
$\eta_{\rm geometry}$ & 0.95(3) & 0.95(3) &  \\
$\eta_{\rm position}$ & 0.95(3) & 0.95(3) &  \\
$\sigma_4$ & 0.189(34) & 0.254(30) & MeV \\
$m$ & 1.2(8) & 0.057(49) &  \\
$c$ & 0.58(16) & 8(20) & MeV \\
$\Delta E$ & 0.100 & 0.100 & MeV \\
\hline
\end{tabular}
\end{center}
\label{tab:parameters}
\end{table}

 The normalization parameters $A_i/\Delta E$ are directly related to the mass $m_i$  of the parent isotope at the top of the chain that would yield the intensity observed if the chain were in equilibrium: 
\begin{eqnarray*}
m_i &=& \frac{A_i}{\Delta E} \frac{1}{\eta St} \sum_j b_jG(\sigma_{ij},m,c) 
\end{eqnarray*}
where $S$ is the activity of the parent isotope,  $t$ is the run time, and $\eta$ is the detection and analysis efficiency, consisting of a product of the geometrical efficiency with which tracks that enter the gas are properly detected at an anode, and the analysis cuts placed on the position and/or energy spectra.

\subsection{EAC Spectral Analysis Results}
\begin{figure}
\includegraphics[width=0.5\textwidth]{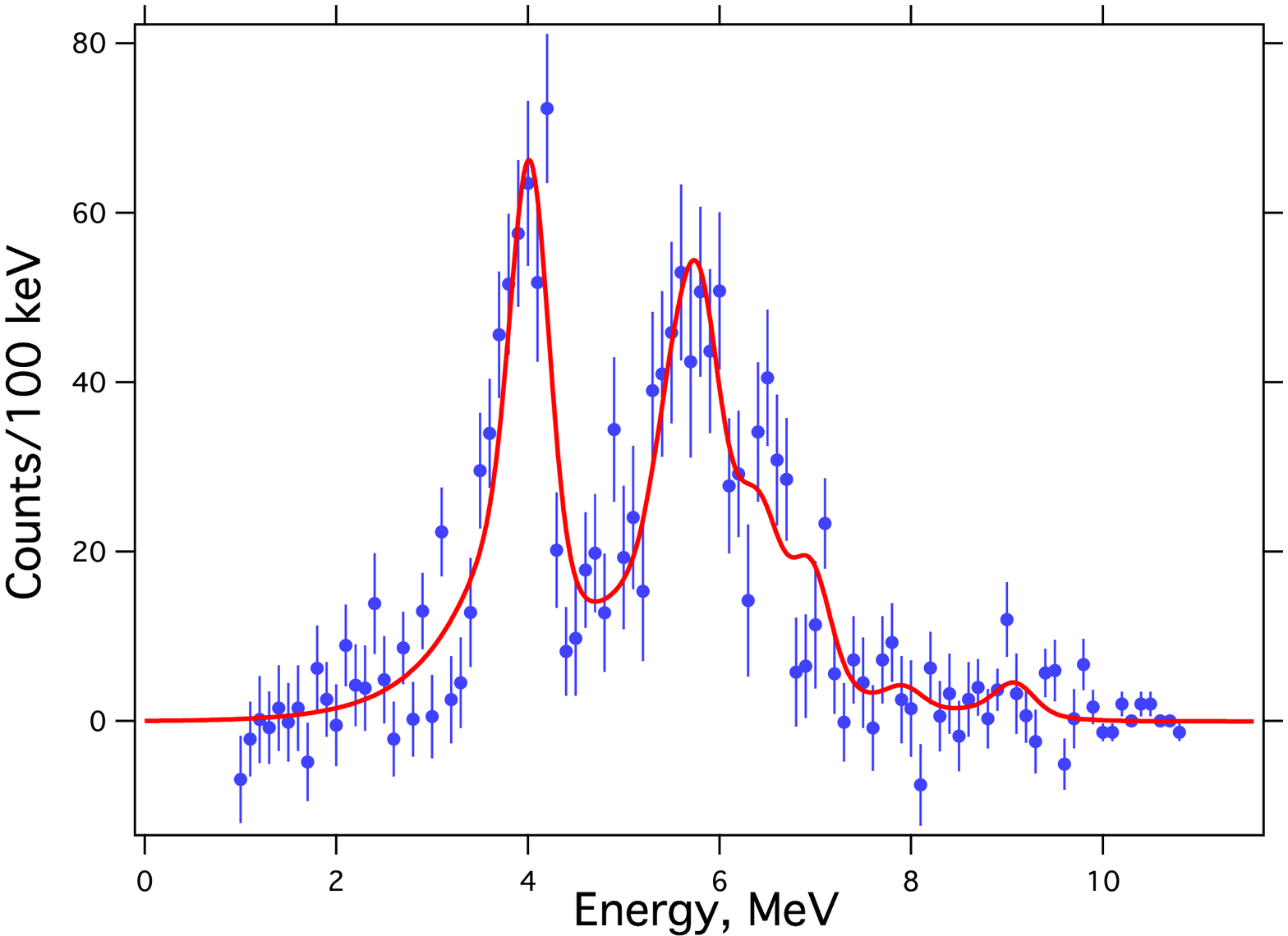}
\includegraphics[width=0.5\textwidth]{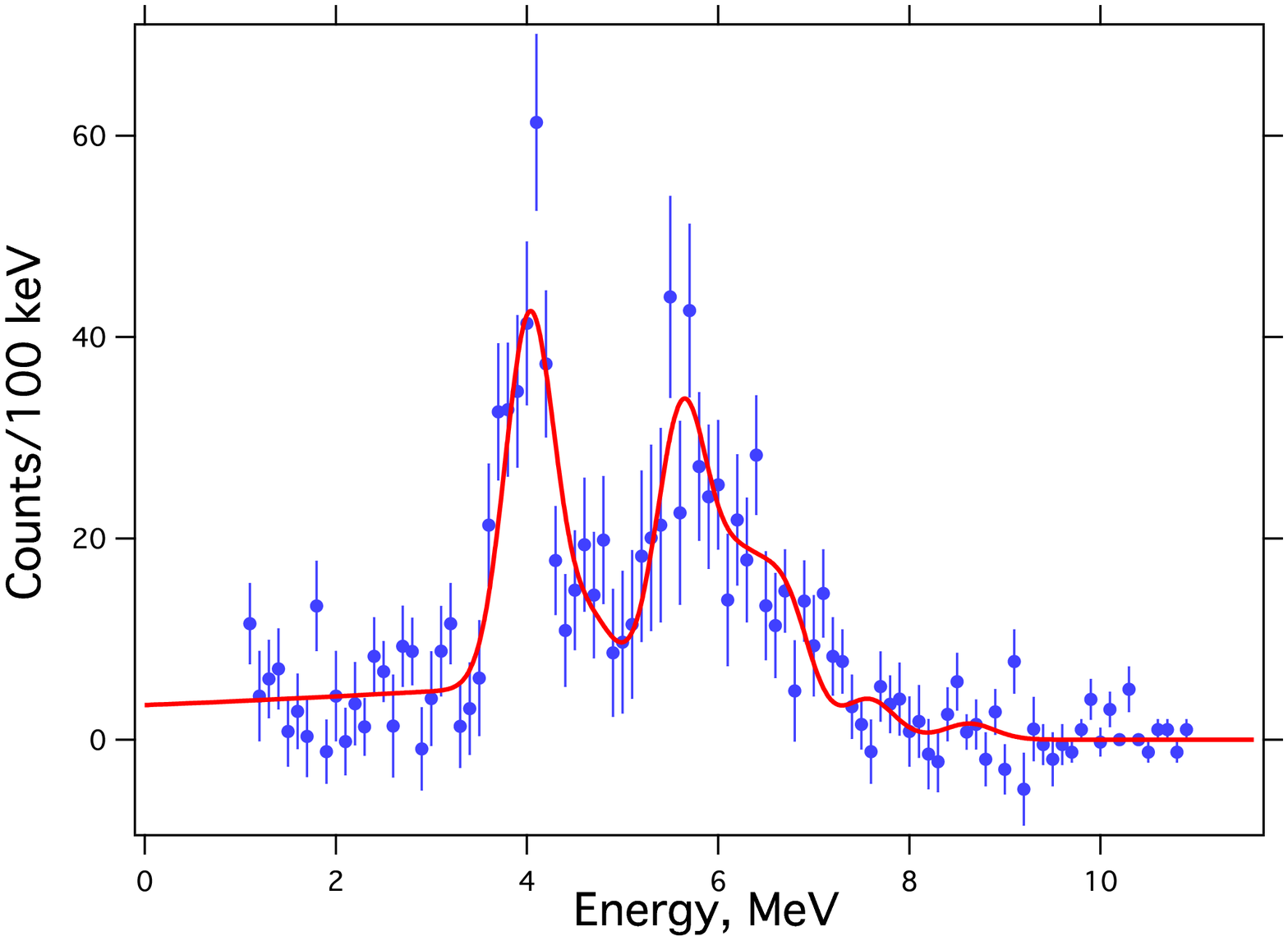}
\begin{center}
\caption{\label{fig:Fit}Background-subtracted spectrum of the K5 upper hotspot (left panel) and lower hotspot (right panel) fit with  uranium- and thorium-chain alpha spectra.  See Table~\ref{tab:chains} for the corresponding activities.}
\end{center}
\end{figure}

The background-subtracted energy spectrum and the best-fit  spectrum of the K5 hotspots are shown in Figure~\ref{fig:Fit}, and the results listed in 
Table~\ref{tab:chains}. 
Both hotspots contain significant amounts of upper and lower thorium-chain isotopes, and barely detectable levels of uranium-chain isotopes.  The other analyses of these hotspots sample only the lower chains.  The individual mass equivalents at each step of the chain can be different owing to disequilibrium, decay, and to loss of activity from the surface caused by alpha recoil.  As discussed below, the difference between $^{232}$Th and $^{228}$Th can be attributed to decay.  Below  $^{228}$Th the half-lives are short and the chain is in secular equilibrium.  Alpha ejection by $^{228}$Th decay may reduce the amount of $^{224}$Ra, which, having a 3-d half-life, will either reattach or be swept away in the gas before decaying.  The daughter $^{220}$Rn may thus be depleted by up to a factor of 2 relative to the surface-fixed situation for $^{224}$Ra.  Ejection of the $^{220}$Rn itself, however, may lead to its decay in the gas since it has a 56-s half-life.  Both the $^{220}$Rn and its daughter, 0.145-s $^{216}$Po, would then be detected with higher efficiency than if they were surface-fixed.  These effects tend to compensate each other, making it difficult to assess whether alpha ejection is an important factor.  However, the last alphas in the chain, $^{212}$Bi and $^{212}$Po, follow decays that must be surface-fixed if they are to be seen, giving a measure of the net loss to alpha ejection.  The low numbers for these alphas suggest  that alpha ejection is playing a significant role, but  statistical precision is limited.  For the purposes of SNO, in which all activities below $^{228}$Th decay in the vessel, only the mass equivalent of that isotope is relevant.   The best estimate of this activity is taken to be the weighted average of the $^{228}$Th, $^{224}$Ra, and $^{220}$Rn results listed in Table \ref{tab:chains}.

\section{EAC Coincidence Analysis}\label{sec:coincAnalysis}
\subsection{Introduction}
Another analysis of the EAC data can give an independent measure of the lower thorium chain activity by counting correlated events that can be attributed to the sequential alpha decays \isotope{Ra}{224} $\rightarrow$ \isotope{Rn}{220} $\rightarrow$ \isotope{Po}{216} $\rightarrow$ \isotope{Pb}{212}.  The energies of the alphas in this series are 5.685\,MeV, 6.288\,MeV and 6.778\,MeV, separated in time according to the 56-s and 0.145-s half-lives of \isotope{Rn}{220} and \isotope{Po}{216}, respectively.   In view of the low efficiency for detection of a complete triple coincidence  given our statistical sample,  we obtain the hotspot strength from the shorter, trailing double-coincidence sample alone.  

\subsection{Event Selection}

\begin{figure}
\begin{center}
\includegraphics[width=0.7\textwidth]{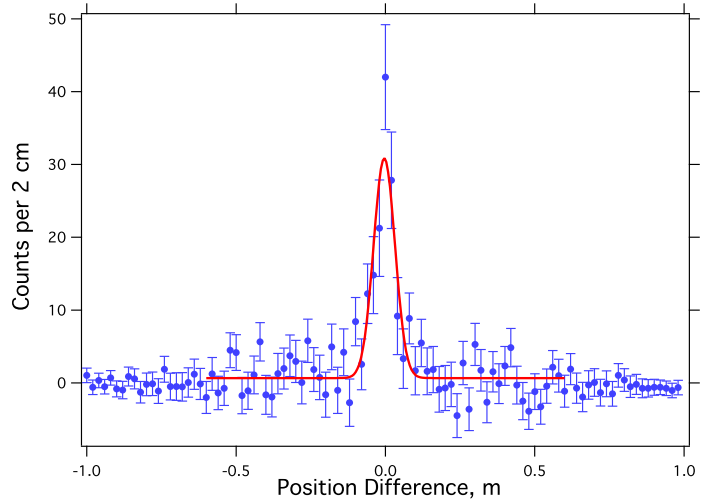}
\end{center}
\caption{The distribution of position differences $\Delta z$ for all coincidence candidate pairs on both K5 hotspots.  Shown are the data with random coincidences subtracted, and the best fit to a Gaussian function.}
\label{fig:zres}
\end{figure}

\begin{figure}
\begin{center}
\includegraphics[width=0.494\textwidth]{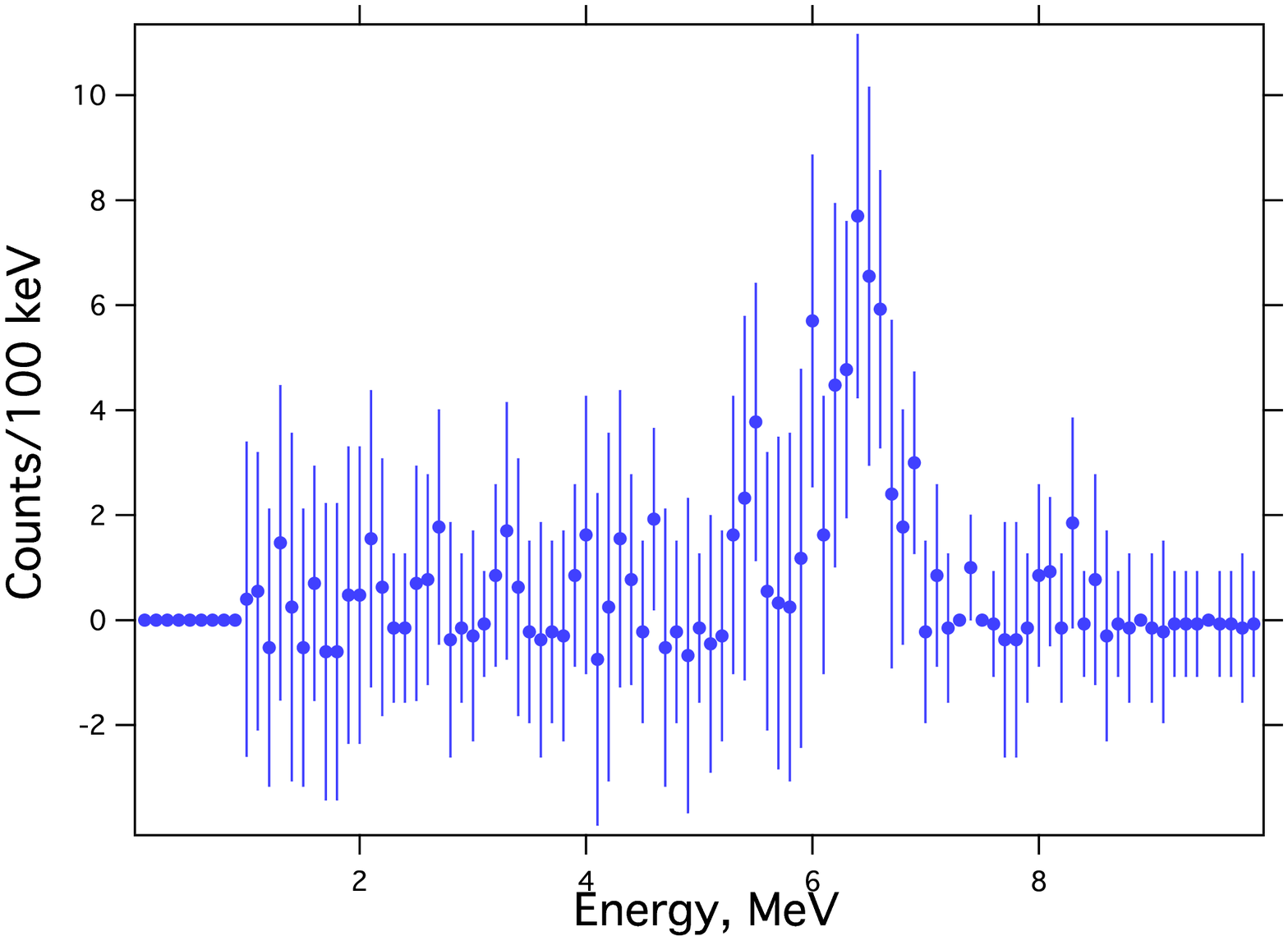}
\includegraphics[width=0.494\textwidth]{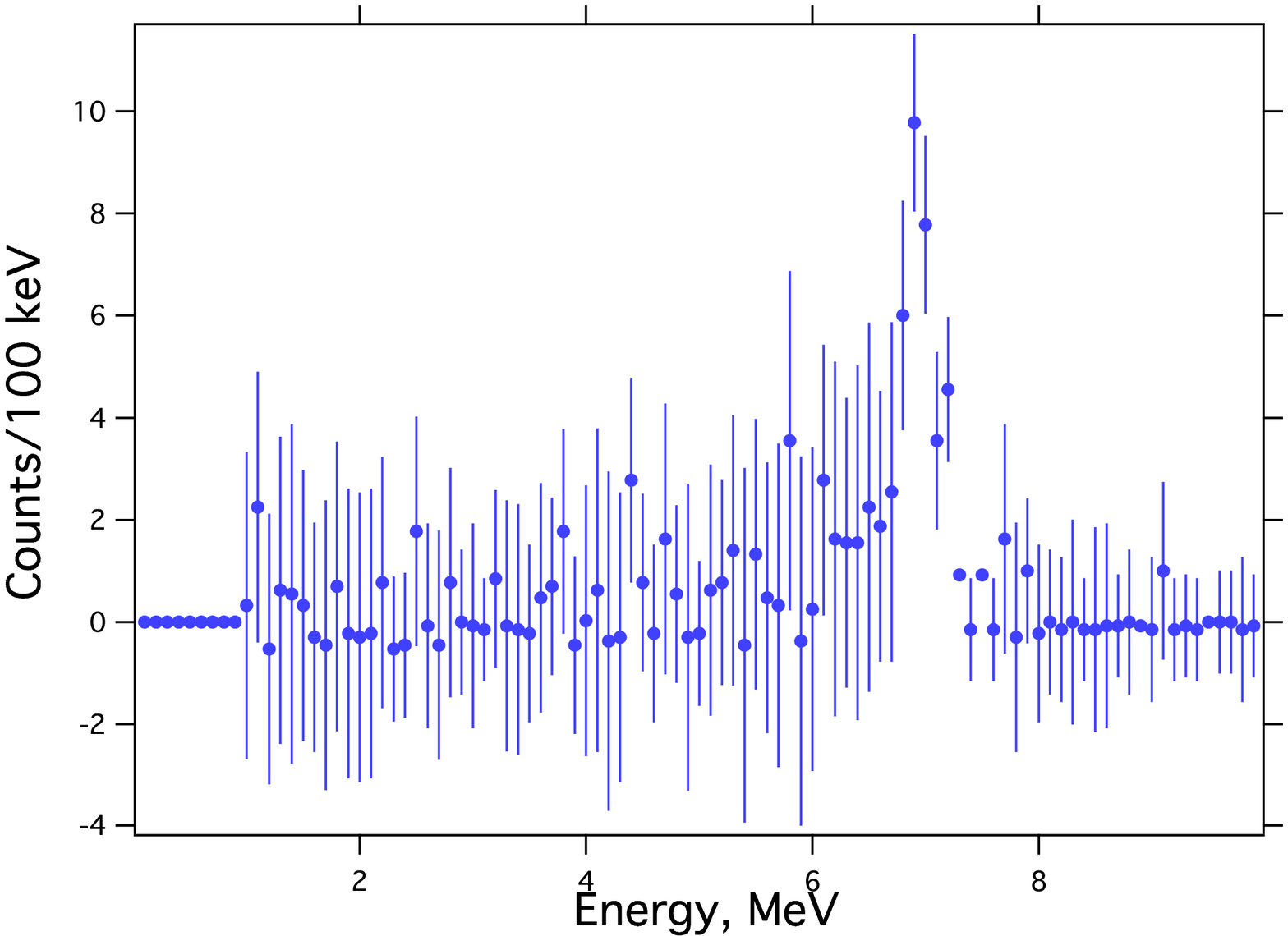} \\
\includegraphics[width=0.494\textwidth]{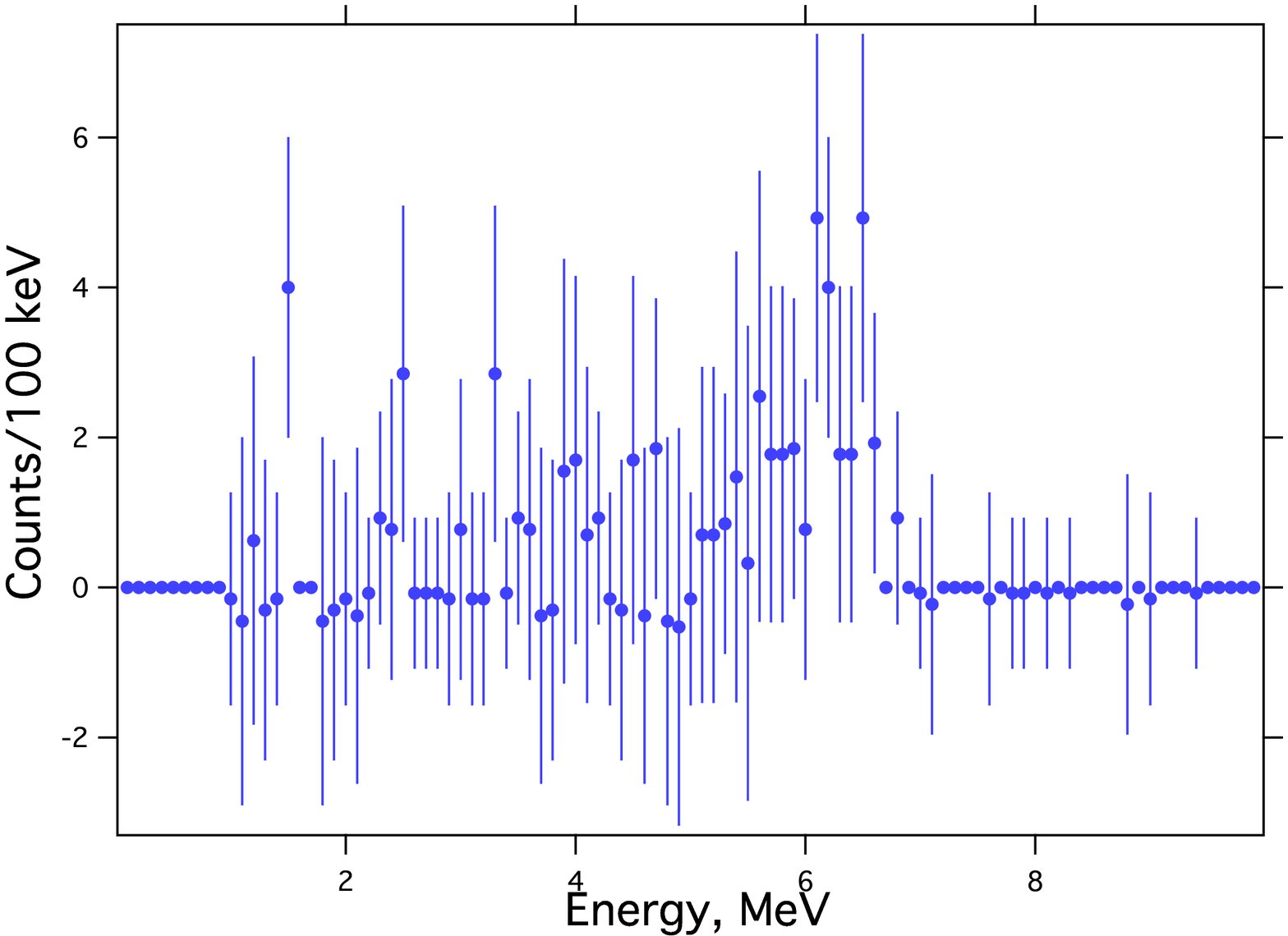}
\includegraphics[width=0.494\textwidth]{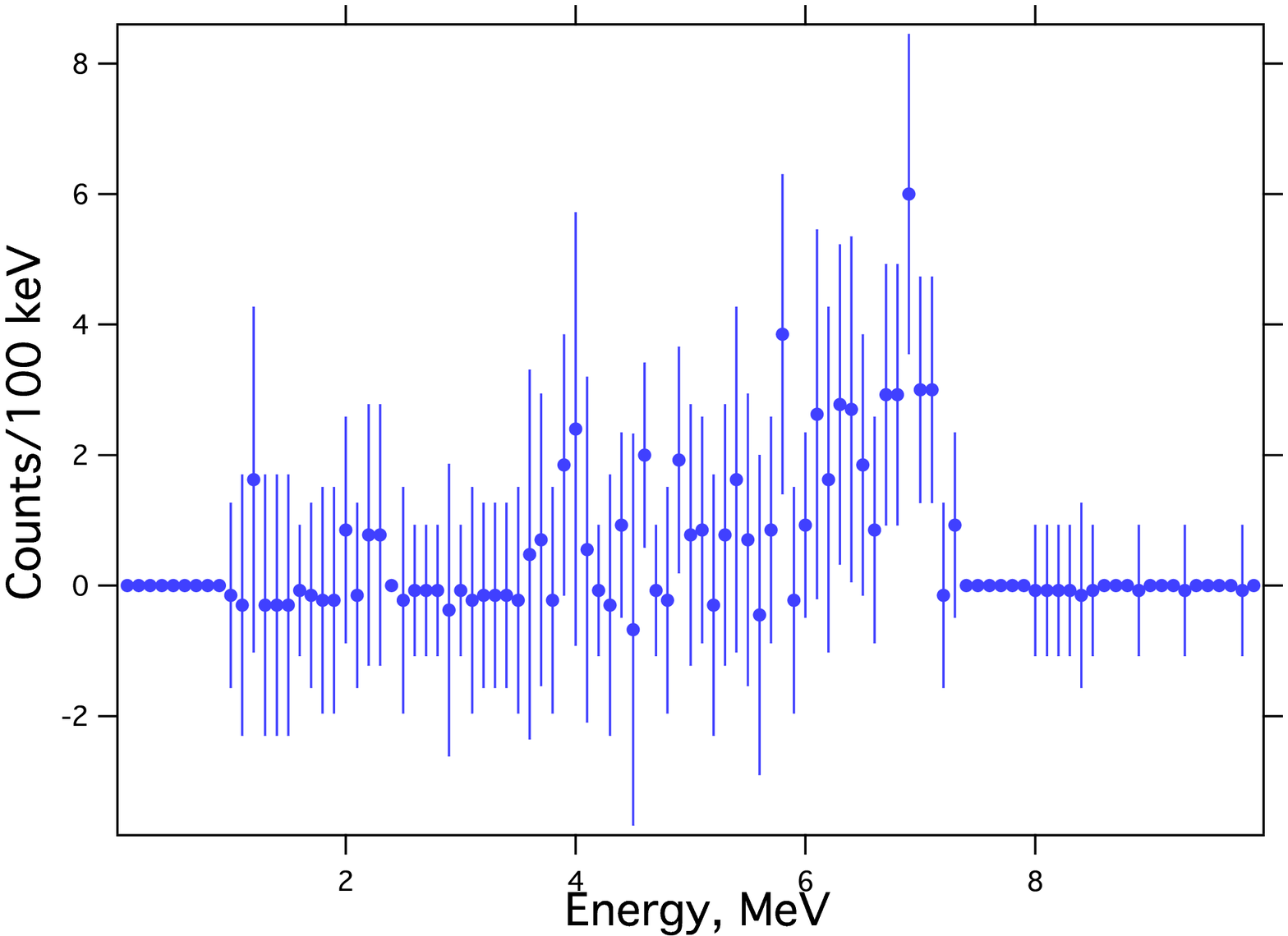}
\end{center}
\caption{The energy spectra of events in double-coincidence candidate pairs from the K5 hotspots: (upper left) the first event on the upper hotspot, (upper right) the second event on the upper hotspot, (lower left) the first event on the lower hotspot, and (lower right) the second event on the lower hotspot.}
\label{fig:coincCandidates}
\end{figure}

\begin{figure}
\begin{center}
\includegraphics[width=0.494\textwidth]{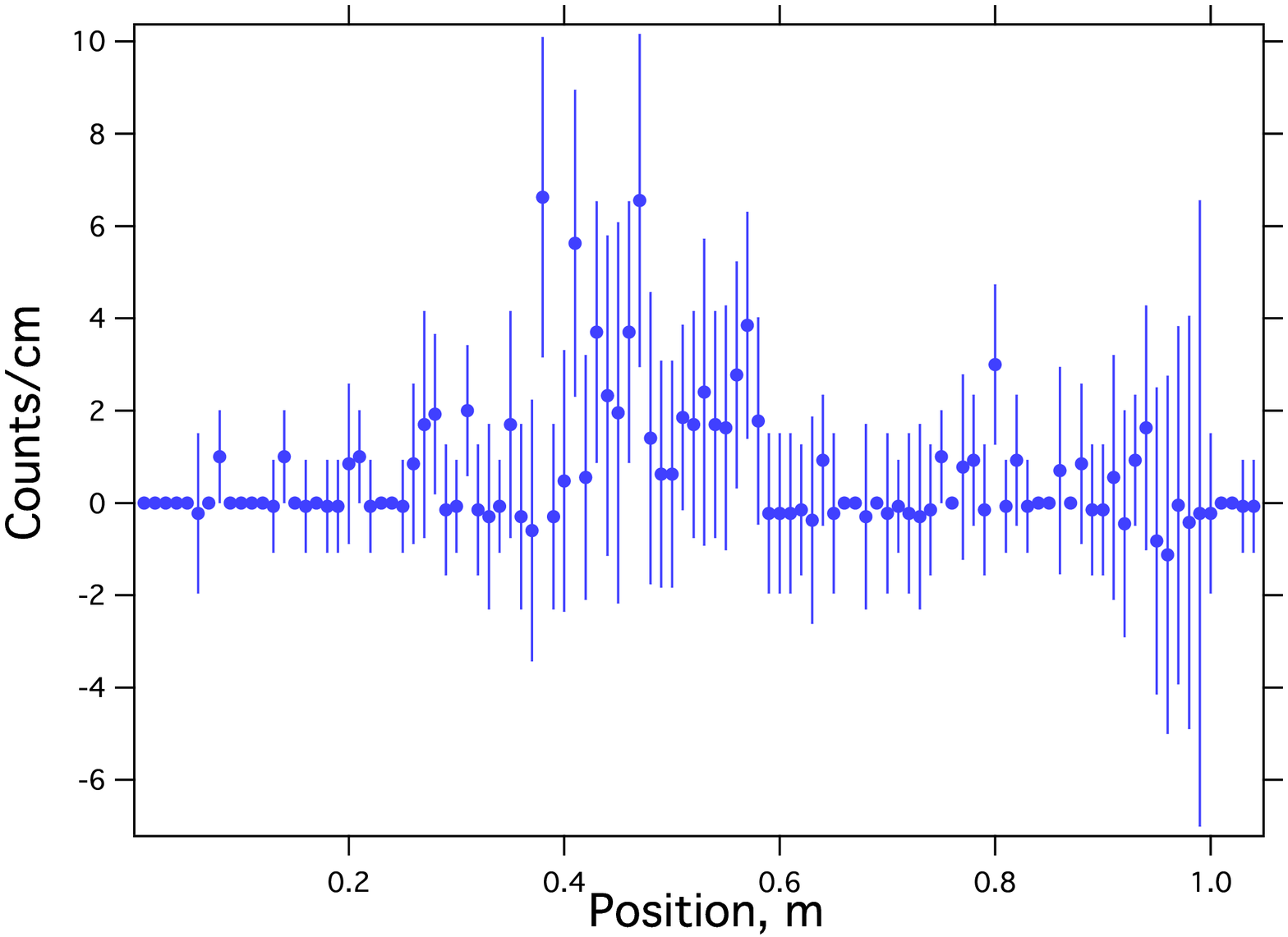}
\includegraphics[width=0.494\textwidth]{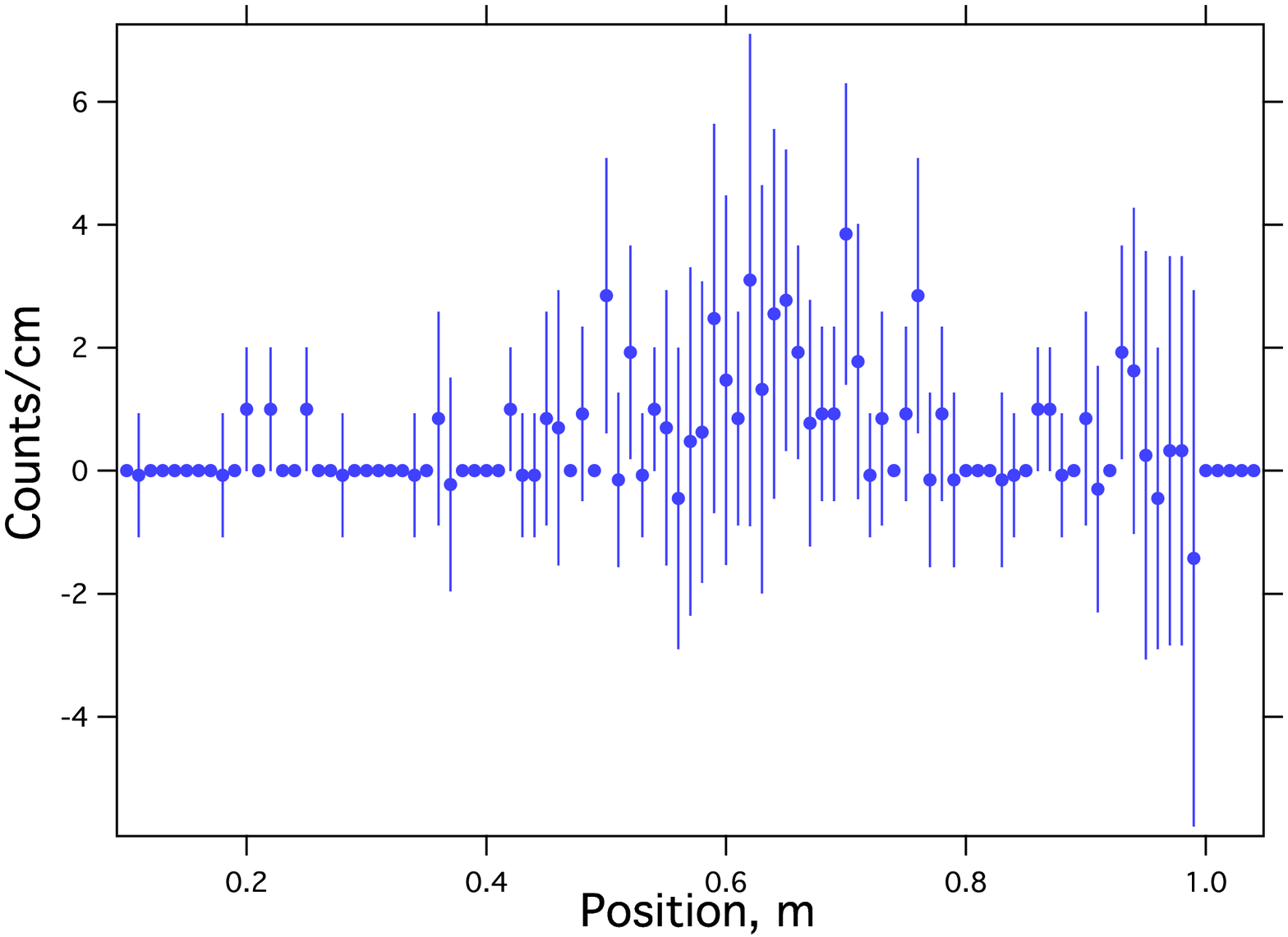}
\end{center}
\caption{The $z$ positions of the first events in double-coincidence candidate pairs from the K5 hotspots: (left panel) the upper hotspot,  (right panel) the  lower hotspot.}
\label{fig:coincPosition}
\end{figure}

Events occurring  with time differences  less than two \isotope{Po}{216} half-lives ($\Delta t \leq  2 t_{\frac{1}{2}} = 0.3$\,sec) are selected as double-coincidence candidates.   Random coincidences and true coincidences with the 56-s half-life are  removed by  renormalizing and subtracting data in a time window from 1 to 5 s.  The instrumental position resolution is obtained from the distribution of differences in position for true-coincidence pairs, as shown in Fig.~\ref{fig:zres}.
The standard deviation $\sigma$ gives the position resolution of the EAC.  The value $\sigma = 3.6 \pm 0.4$\,cm is the spread in $\Delta z$ measured for time-correlated pairs of events.  The intrinsic single-event resolution is therefore $\sigma/\sqrt{2} = 2.5 \pm 0.3$\,cm.  Figure~\ref{fig:coincCandidates} shows the subtracted energy spectra of both alphas in pairs satisfying the time condition and having $\left| \Delta z \right| \leq 5$\,cm.
 Figure~\ref{fig:coincPosition} shows the position distributions for the first  events in double-coincidence candidate pairs.
The positions of the coincidence events are consistent  with the $z$-distributions in the EAC spectral analysis data set (Fig.~\ref{fig:Z}).     The statistics of the two K5 hotspots are summarized in Table~\ref{tab:coincCuts}.
\begin{table}
\caption{Statistics of  the EAC coincidence analysis of the two K5 hotspot data sets.}
\label{tab:coincCuts}
\begin{center}
\begin{tabular}{r|cc}
\hline \hline
 & K5 upper & K5 lower \\ 
\hline
live time (s)          & 105\,844 & 95\,578 \\ 
total events            & 7778        & 4330      \\ 
double candidates   & 96      & 69  \\
corrected for randoms   & 68.9      & 55.0    \\
\hline \hline
\end{tabular}
\end{center}
\end{table}

\subsection{Double Coincidence Detection Efficiency}

The EAC detection efficiency for double coincidences depends on the extent to which the thorium-chain daughters are surface-fixed.  In particular, if the $^{220}$Rn is ejected into the gas (and not subsequently swept or diffused away), the efficiency for detecting the coincidence decay may be essentially unity.  This could occur for  half the $^{224}$Ra decays, i.e. when the alpha is directed into the metal.  For the other half, the double coincidences can be expected to be seen 25\% of the time.  The total doubles efficiency is then 62.5\%.  If, on the other hand, the $^{220}$Rn remains surface-fixed in all $^{224}$Ra decays, the doubles efficiency is 25\%. Both of these efficiencies are further reduced by the time window (the 0.3-s time window includes 76.1\% of decays of 0.145-s  $^{216}$Po), by the geometrical efficiency of the wires (0.95 for each event in the double), and by the position cut ($0.84\pm0.05$ for the combined coincidence event).  For surface-fixed $^{220}$Rn, the efficiency to see a double coincidence pair for a given traversal of the lower thorium chain is 14.4\%.    If the $^{220}$Rn were being ejected but there were no corresponding ejection process for the parent $^{224}$Ra, the efficiency would be 2.5 times larger. The results are summarized in Table~\ref{tab:coincResults}.  
\begin{table}[ht]
\caption{\label{tab:coincResults} Mass of thorium measured by the EAC coincidence method.  The asymmetrical uncertainties reflect the range of detection efficiencies.}
\begin{center}
\begin{tabular}{lcc}
\hline \hline
Hotspot      & Events            & $m_{\rm Th} (\mu$g) \\
\hline
\vspace{-0.1in} & \\
K5 Upper   & 68.9              & $1.11^{+ 0.17}_{-0.66}$ \\
\vspace{-0.1in} & \\
K5 Lower    & 55.0             & $0.98^{+ 0.16}_{-0.59}$ \\
\vspace{-0.1in} & \\
\hline \hline
\end{tabular}
\end{center}
\end{table}

\section{Acid Elution}

\subsection{Introduction}

		From the EAC data it was found that both hotspots are primarily thorium-chain isotopes, on the surface, and located at the positions  suggested by the original Cherenkov analysis.  Any thorium or radium daughters on the nickel surface of the NCD are easily removed by rinsing with weak (0.1\,mol/L) hydrochloric acid (HCl).  The method of eluting the contaminated NCD is based upon the techniques used in hydrous titanium oxide (HTiO) assays, which monitored the $^{232}$Th and $^{238}$U content of the D$_2$O and H$_2$O throughout the life of the SNO experiment~\cite{snoHTiO_03,snoHTiO_08}.  We discuss here only the methods particular to the elution of a section of an NCD.   The radioactive content of the acid eluate can be determined by concentrating the solution to a volume of a few mL, mixing it with liquid scintillator and counting the sample using a beta-alpha coincidence method.  Beta-alpha coincidences exist in the lower parts of both the uranium and thorium chains, and the time intervals between the initial betas and subsequent alphas are sufficiently different that discrimination between the two chains is possible.  

\subsection{Assay of an NCD Section}
\begin{figure}
\begin{center}
\includegraphics[width=0.35\textwidth]{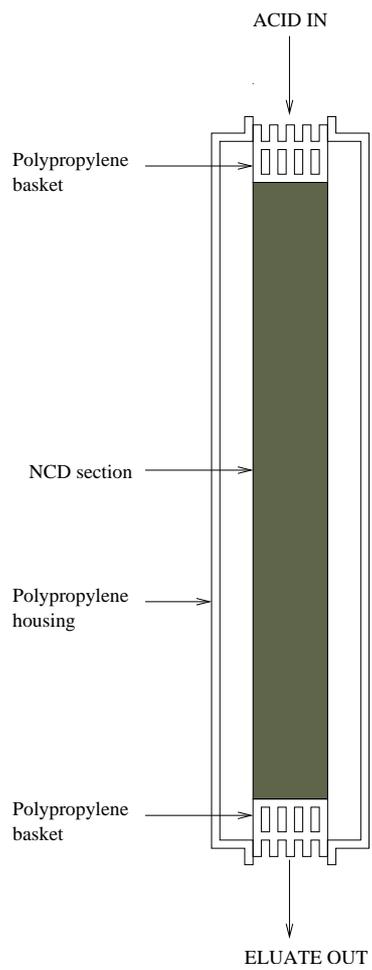}
\end{center}
\caption{A section of the NCD nickel tube mounted in the elution rig, showing the polypropylene baskets at the ends.}
\label{fig:elutionbaskets}
\end{figure}

The assay of an NCD section consists of three stages: elution, secondary concentration and counting.  Specially constructed polypropylene baskets are placed on each end of the NCD section (Fig.~\ref{fig:elutionbaskets}).  
Compression of an O-ring at the interface between the NCD and a basket ensures only the outer surface of the NCD section is eluted with acid.  This assembly is placed inside a polypropylene column  that is connected to the elution rig~\cite{snoHTiO_03}.   The elution rig holds 25-cm or 100-cm columns, which restrict the length of NCD
section to approximately 14\,cm or 90\,cm, respectively.  The elution rig removes thorium and radium isotopes from the surface of an NCD section by circulating 0.1\,mol/L HCl through the columns at a flow rate of approximately 80\,L/min for 20 minutes.  The 15\,L of acid eluate is then concentrated by extracting heavy isotopes onto a cation exchange resin (Dowex 50WX8, H$^{+}$ form, 100 mesh).  Radium is selectively eluted from the resin using 0.25\,mol/L EDTA (ethylene diamine tetraacetic acid).  The solution is further concentrated by boiling with concentrated HNO$_3$ and co-precipitation of radium with HTiO.  Details of the chemistry and methods are described in \cite{snoHTiO_08}.  

The final sample is added to 42\,g of OptiPhase HiSafe-3 liquid scintillator and sealed.  The sealed sample is optically coupled to a 5-cm diameter PMT connected to a NIM-based electronics system where a series of logic conditions is applied to the incoming pulses to determine when a genuine coincidence has occurred.  Samples are counted for 10--14 days.  The counting efficiencies are measured to be $45 \pm 5$\,\% for \isotope{Ra}{224} coincidences and $60 \pm 10$\,\% for \isotope{Ra}{226} coincidences.  The counting system is fully described in \cite{RKT}.

  The secondary concentration methods described in \cite{snoHTiO_08} have been tested using a nickel concentration of $\sim$9\,ppm in the 15\,L eluate.  Concentrations up to this level are found to have no effect on the efficiency.   To determine the concentration of nickel arising from elution of an NCD, a 14-cm section of a damaged but radiopure NCD was eluted for 20 minutes in the manner described above.  A sample of the HCl was taken before and after elution and sent for Inductively Coupled Plasma Mass Spectrometry (ICP-MS).  Assuming the amount of nickel removed from an NCD section during elution is directly proportional to the length of the section, a 90-cm section would result in a nickel concentration of 4.3\,ppm in the 15\,L eluate, well below the limit of 9\,ppm.

\subsection{Validation of the Elution Method}
To ensure that the Acid-elution method can be used to give an accurate measure of radioactive contamination on the surface of an NCD section, a series of spike experiments were performed by contaminating radiopure NCDs with \isotope{Th}{228} \cite{HMOK}.  The capped NCDs were soaked in a weak $^{228}$Th solution for time periods ranging from 1--7 days.  During this time, thorium plated onto the surface of the NCD where it remained after the NCD was removed from solution.  The activity remaining in the solution was determined using HTiO co-precipitation and beta-alpha coincidence counting.  Following a correction for plating onto the end caps and container, the difference in activity between the initial and final solutions was used to determine how much thorium plated onto the NCD section.

Before the spiked NCD section was eluted, the three-stage assay process was carried out using an uncontaminated section of NCD to determine the equipment and reagent background.  Before cutting test NCD sections, a pipe cutter was ultrasonically cleaned with Radiac solution and handled at all times using gloves.  The cut NCD section with baskets was placed inside a polypropylene column, eluted, concentrated and counted.  The measured count rate from the spike elution was corrected for the background count rate.  The amount of $^{224}$Ra present on the spiked NCD section before elution, is given by  
\begin{displaymath}
A_{\rm soe} =\frac{ \big(N_{\rm elu} - N_{\rm bkg}e^{-\lambda_{\rm Ra224}\Delta
t_{\rm bs}}\big)}{\epsilon_{\rm count} e^{-\lambda_{\rm Ra224} t_{\rm el}}},
\end{displaymath}
where $N_{\rm elu}$ is the measured count rate from the elution of the spiked NCD, $N_{\rm bkg}$ is the measured count rate from the background sample, $\Delta t_{\rm bs}$ is the difference in time between the start of elution and the start of counting for the background and spiked samples, $\epsilon_{\rm count}$ is the counting efficiency and $t_{\rm el}$ was the time taken for the elution and secondary concentration of the spiked sample.  The calculated activity at the start of elution was compared with the amount of thorium that plated onto the NCD to obtain the (combined) elution and secondary concentration efficiency.

In a total of seven spike tests, the mean efficiency of the elution and secondary concentration stages was found to be $45 \pm 18$\,\%,  in good agreement with that for the applicable stages of the HTiO assay method ($56 \pm 6$\,\%).  A large proportion of the uncertainty quoted for these efficiencies is due to uncertainties in the measurement of
the solution activity after soaking.  This error is not included in the measurement of hotspot activity, so it is reasonable to use the elution and secondary concentration efficiencies of the HTiO assay technique.  The combined elution, secondary concentration, and counting efficiencies give overall efficiencies of $31 \pm 7$\,\% for $^{226}$Ra and $23 \pm 4$\,\% for $^{224}$Ra. 

\subsection{Elution of K5}

Before elution of the K5 hotspot, the background contribution from equipment and reagents was measured using an uncontaminated 90-cm section of K5.  Following this, the elution rig was cleaned by circulating 0.5\,mol/L HCl, for approximately 40 minutes, then rinsed with ultra-pure water until a near-neutral pH was reached.  The background measurement was performed approximately 5 days before the hotspot elution and the results were consistent with previous measurements made using the HTiO technique \cite{HMOK}.  Due to restrictions on the size of columns that could be connected to the elution rig, K5 was cut into two sections, informed by the EAC results.    The sections were eluted, concentrated and counted separately.   The results are included below in Table \ref{tab:combinedResults}.

\section{Conclusions}

We have applied four methods to determine the strengths of two radioactive hotspots on one of SNO's NCDs.  Cherenkov light from radioactivity was analyzed using methods developed for general Cherenkov analysis in SNO.  A position-sensitive multi-wire proportional counter that surrounded the contaminated portions of the NCD surface provided a detailed alpha spectrum.  Alpha decay events were analyzed according to their energy spectra in one method, while a second method exploited the time-correlated nature of these decays.  In the fourth method  activity from the positions indicated by the first two methods was removed and subjected to radiochemical assay to derive the hotspot activities.  Where results can be compared, the laboratory methods are generally in agreement   (see Table~\ref{tab:combinedResults}).

\begin{table}
\caption{\label{tab:combinedResults} A summary of the results for K5 and K2 hotspot strengths, measured by the mass of the parent nuclei in $\mu$g, uncorrected for decay (see text).  The results for the EAC spectral analysis consist of weighted averages of the $^{228}$Th, $^{224}$Ra, and $^{220}$Rn   groups for the lower thorium chain, and $^{218}$Po for the lower uranium chain.  The upper chains are represented by $^{232}$Th and $^{230}$Th.}
\begin{center}
\begin{tabular}{ll|ccccc}
\hline \hline
Hotspot   & Method       & $m_{\rm Th}^{\rm up}$ & $m_{\rm Th}^{\rm lo}$ & $m_{\rm U}^{\rm up}$ & $m_{\rm U}^{\rm lo}$ \\
\hline
K5 Upper  & Spectral       & 2.65$\pm$0.69      & 0.81$\pm$0.19      & 0.01$\pm$0.07  & 0.05$\pm$0.04   \\
\vspace{-0.1in} & \\
               & Coincidence & n/a                          & $1.11^{+ 0.17}_{-0.66}$      & n/a                         & n/a \\
\vspace{-0.1in} & \\
               & Elution         & n/a                          & 0.61$\pm$0.17      & n/a                         & $<$0.07 \\
\hline
K5 Lower & Spectral       &  1.83$\pm$0.22       &  0.61$\pm$0.09      & 0.12$\pm$0.06              & 0.06$\pm$0.02 \\
\vspace{-0.1in} & \\
               & Coincidence & n/a                          & $0.98^{+ 0.16}_{-0.59}$      & n/a                         & n/a                     \\
\vspace{-0.1in} & \\
               & Elution         & n/a                          & 0.30$\pm$0.12      & n/a                         & 0.11$\pm$0.03 \\
\hline
\vspace{-0.1in} & \\
K5 total  & Cherenkov      & n/a     & $1.48_{-0.27}^{+0.24}$      & n/a             & $0.77_{-0.23}^{+0.19}$  \\
\vspace{-0.1in} & \\
  & Cherenkov     & n/a     & $1.69_{-0.28}^{+0.31}$      & n/a             & $<  0.20$  \\
& constrained \\
\hline
K2 & Ge Detector & n/a & $1.43\pm 0.17$ & n/a & $<0.40$ \\
\hline
K2 & Cherenkov & n/a & $<0.93$ & n/a & $\equiv 0$  \\
& constrained \\
\hline \hline
\end{tabular}
\end{center}
\end{table}

Further understanding of the results can be obtained from the chronology.  The time at which the hotspots were deposited on the NCD surfaces can be pinned down quite well because no activity was seen in Cherenkov light before diagnostic work was carried out on K5 in April 2004.  The data point to a selective chemistry that deposited Th isotopes on the nickel surface at that time.  Solar neutrino measurements \cite{ncdprl} took place between November 27, 2004 and November 28, 2006, and the laboratory measurements on the hotspots in May 2007.  Neglecting ingrowth and decay of 5.75-y $^{228}$Ra,  the $^{228}$Th deposited would have decayed with its 1.91-y half-life, to reach an activity at the time of measurement 0.327 times its initial level.  Taking into account  this decay, the mass equivalents of  $^{232}$Th and $^{228}$Th are in agreement, indicating the chain was in equilibrium at the time of deposit.  

The Cherenkov analysis integrates over the two-year period beginning November 27, 2004, and so should report an average lower-thorium-chain activity that is 0.56 times the initial level, and 1.71 times the level measured subsequently in the laboratory.   With the latter figure, it can be seen that the Cherenkov analysis yields thorium-chain activity quite consistent with the other methods,  but  larger uranium-chain activity.  Distinguishing U from Th is based on the isotropy parameter, the extraction of which for surface deposits and low-energy radiation is more difficult than in its usual application.      The very low levels of uranium-chain activity in the laboratory data can be used to constrain the Cherenkov fit and thereby obviate the need to rely on the isotropy parameter.   There is additional evidence that the U levels are extremely small.  The Th chain alpha data imply that $^{228}$Ra is absent and not supporting $^{228}$Th.  Therefore $^{226}$Ra must also be essentially absent, and without it, the rest of the lower U chain. The U chain is further limited by the absence of $^{230}$Th, which shows the U chain was not significantly present in the original source from which the deposits were laid down.  Subsequent leaching of $^{226}$Ra then even further discriminated against the lower U chain.

The K5 lower hotspot shows small positive numbers for the lower U chain, but even this amount  may simply reflect imperfect correction for background $^{222}$Rn, which is ubiquitous.   We therefore apply a constraint of zero U-chain activity to the Cherenkov analysis, but enlarge the uncertainty to encompass the results from the laboratory data.  The result of this constraint (expressed for the same time interval), $1.69_{-0.28}^{+0.31}$ $\mu$g equivalent $^{232}$Th, is shown in Table~\ref{tab:combinedResults}, and is the value recommended for use in future analyses of the third phase of the SNO experiment.  The unconstrained values were used in \cite{ncdprl}.

We omit discussion of the K2 hotspot, which was analyzed in similar ways, and also by direct counting with a low-background Ge detector in the SNO laboratory.  In contrast to the K5 hotspot, the K2 hotspot showed no evidence of disequilibrium and very little surface activity.  It is evidently an inclusion within the material. The results are included in Table~\ref{tab:combinedResults}.  Germanium-detector counting could have been applied to K5 as well, of course, but the time required for it was prohibitive.  Instead, the Acid-elution method on K5 was carried out in parallel with K2 counting. 

The results of this work reduced a potential systematic uncertainty of 4\,\% due to unknown background sources in the total \isotope{B}{8}  solar neutrino flux measured in Phase III of the SNO experiment to less than 1\,\%.  The contamination was restricted to two regions of a single NCD totaling about 40\,cm in length (see Figure \ref{fig:Z}).  This is only about 0.1\,\% of the total length of the NCD array, but the consequences could have been disproportionately adverse to SNO.   It is not known how K5 became contaminated so we can draw no conclusions about how the contamination could have been prevented.  The amount of \isotope{Th}{232} determined to exist on both hotspots is quite large given known concentrations in typical materials.  The obvious hypothesis that the contamination was due to intrusion of dust from the surrounding mine is ruled out, since it would require several grams of dust to get the observed amount of thorium-chain contamination.  Such a quantity of dust would be plainly visible on the surface of K5 when in fact the hotspot regions were utterly indistinguishable in appearance from a typical uncontaminated NCD section.   

The methods described here are sensitive to uranium- and thorium-chain contamination at the level of tens of ng.  While one might expect similar sized contributions from both the uranium and thorium chains, the chemistry of these elements is quite different and thus contamination will depend on manufacturing processes, storage and handling conditions etc.  Current and future generations of neutrino and dark matter detection experiments have even more stringent radiopurity requirements than SNO, and it should be expected that at least some of these experiments will experience similar unexpected hotspots.  The methods described here can help diagnose and mitigate the effects.  

\section{Acknowledgments}
This work was supported by the United States Department of Energy  Office of Science Division of Nuclear Physics, the United Kingdom Science and Technology Facilities Council (formerly PPARC) and the Natural Sciences and Engineering Research Council of Canada.  We thank the SNO technical staff for assistance with this work and also Vale-INCO for hosting our underground facilities.  Critical mechanical components of the equipment for alpha counting and acid elution were fabricated at the CENPA Machine Shop at the University of Washington and the Physics Mechanical Workshop at the University of Oxford, respectively.  The ICP-MS analysis of HCl samples was performed at Geolabs, Sudbury, ON.


\bibliographystyle{unsrt}
\bibliography{sources}

\end{document}